\documentclass[10pt,twocolumn,letterpaper]{article}

\usepackage[pagenumbers]{cvpr}

\usepackage{amsmath}
\usepackage{amsthm} %
\usepackage{amssymb}

\usepackage{graphicx}
\usepackage[export]{adjustbox}
\usepackage{float}
\usepackage{stfloats}
\usepackage{booktabs}  %
\usepackage[list=true]{subcaption}
\usepackage{longtable}
\usepackage{multirow}
\usepackage[table, x11names]{xcolor}
\usepackage{rotating}
\usepackage{makecell}
\usepackage{stfloats}
\newcommand{\centered}[1]{\begin{tabular}{l} #1 \end{tabular}}

\makeatletter %
\@namedef{ver@everyshi.sty}{}
\makeatother
\usepackage{pgfplots}  %
\usetikzlibrary{pgfplots.groupplots}
\usepgfplotslibrary{units}
\usetikzlibrary{positioning}
\pgfplotsset{compat=newest}
\usetikzlibrary{spy}
\usetikzlibrary{matrix}

\newtheorem{thm}{Theorem}

\usepackage[createShortEnv]{proof-at-the-end}

\usepackage{listings}

\definecolor{codegreen}{rgb}{0,0.6,0}
\definecolor{codegray}{rgb}{0.5,0.5,0.5}
\definecolor{codepurple}{rgb}{0.58,0,0.82}
\definecolor{backcolour}{rgb}{0.95,0.95,0.92}

\lstdefinestyle{mystyle}{
    backgroundcolor=\color{backcolour},   
    commentstyle=\color{codegreen},
    keywordstyle=\color{magenta},
    numberstyle=\tiny\color{codegray},
    stringstyle=\color{codepurple},
    basicstyle=\ttfamily\footnotesize,
    breakatwhitespace=false,         
    breaklines=true,                 
    captionpos=b,                    
    keepspaces=true,                 
    numbers=left,                    
    numbersep=5pt,                  
    showspaces=false,                
    showstringspaces=false,
    showtabs=false,                  
    tabsize=2
}

\usepackage[pagebackref,breaklinks,colorlinks]{hyperref}

\usepackage{cite}

\usepackage[capitalize, nameinlink]{cleveref}
\Crefname{subsection}{Subsection}{Subsections}

\usepackage{enumitem}
\usepackage[shortcuts]{extdash}

\DeclareMathOperator{\dct}{\text{2D-DCT}}
\DeclareMathOperator{\var}{\text{Var}}
\DeclareMathOperator{\vgg}{\text{VGG}_{5,4}}
\DeclareMathOperator{\jpeg}{\textit{JPEG}}
\newcommand{\norm}[1]{\left\lVert#1\right\rVert}
\newcommand{\round}[1]{\ensuremath{\lfloor#1\rceil}}

\newlength{\ffhqw}
\newlength{\ffhql}
\newcommand{\real}[4][\ffhqw]{\centered{\includegraphics[width=#1]{images/ours/#2-#3/real/#4.jpg}}}

\newcommand{\noisy}[4][\ffhqw]{\centered{\includegraphics[width=#1]{images/ours/#2-#3/compressed/#4.jpg}}}

\newcommand{\mmse}[4][\ffhqw]{\centered{\includegraphics[width=#1]{images/mmse/#2-#3/fake_0/#4.jpg}}}

\newcommand{\pscgan}[5][\ffhqw]{\centered{\includegraphics[width=#1]{images/pscgan/#2-#3/fake_#5/#4.jpg}}}
\newcommand{\pscganc}[5][\ffhqw]{\centered{\includegraphics[width=#1]{images/pscgan/#2-#3/fake_compressed_#5/#4.jpg}}}
\newcommand{\pscgand}[5][\ffhqw]{\centered{\includegraphics[width=#1]{images/pscgan/#2-#3/diff_#5/#4.jpg}}}
\newcommand{\pscgans}[4][\ffhqw]{\centered{\includegraphics[width=#1]{images/pscgan/#2-#3/std/#4.jpg}}}

\newcommand{\pscganp}[5][\ffhqw]{\centered{\includegraphics[width=#1]{images/pscgan-p/#2-#3/fake_#5/#4.jpg}}}

\newcommand{\ours}[5][\ffhqw]{\centered{\includegraphics[width=#1]{images/ours/#2-#3/fake_#5/#4.jpg}}}
\newcommand{\oursc}[5][\ffhqw]{\centered{\includegraphics[width=#1]{images/ours/#2-#3/fake_compressed_#5/#4.jpg}}}
\newcommand{\oursd}[5][\ffhqw]{\centered{\includegraphics[width=#1]{images/ours/#2-#3/diff_#5/#4.jpg}}}
\newcommand{\ourss}[4][\ffhqw]{\centered{\includegraphics[width=#1]{images/ours/#2-#3/std/#4.jpg}}}
\newcommand{\oursa}[4][\ffhqw]{\centered{\includegraphics[width=#1]{images/ours/#2-#3/mean/#4.jpg}}}

\newcommand{\oursp}[5][\ffhqw]{\centered{\includegraphics[width=#1]{images/ours-p/#2-#3/fake_#5/#4.jpg}}}

\newcommand{\ourspc}[5][\ffhqw]{\centered{\includegraphics[width=#1]{images/ours-p/#2-#3/fake_compressed_#5/#4.jpg}}}
\newcommand{\ourspd}[5][\ffhqw]{\centered{\includegraphics[width=#1]{images/ours-p/#2-#3/diff_#5/#4.jpg}}}

\newcommand{\bahat}[5][\ffhqw]{\centered{\includegraphics[width=#1]{images/bahat/#2-#3/fake_#5/#4.jpg}}}
\newcommand{\bahats}[4][\ffhqw]{\centered{\includegraphics[width=#1]{images/bahat/#2-#3/std/#4.jpg}}}

\newcommand{\qgac}[4][\ffhqw]{\centered{\includegraphics[width=#1]{images/qgac/#2-#3/#4.jpg}}}
\newcommand{\qgacg}[4][\ffhqw]{\centered{\includegraphics[width=#1]{images/qgac-gan/#2-#3/#4.jpg}}}

\def\cvprPaperID{37} %
\def\confName{NTIRE}
\def\confYear{2023}

\begin{document}

\title{High-Perceptual Quality JPEG Decoding via Posterior Sampling}

\author{
	Sean Man \qquad Guy Ohayon \qquad Theo Adrai \qquad Michael Elad \\
	Technion - Israel Institute of Technology, Haifa, Israel \\
	{\tt\small \{sean.man,ohayonguy,theoad\}@campus.technion.ac.il, elad@cs.technion.ac.il}
}

\maketitle

\begin{abstract}

JPEG is arguably the most popular image coding format, achieving high compression ratios via lossy quantization that may create visual artifacts degradation. Numerous attempts to remove these artifacts were conceived over the years, and common to most of these is the use of deterministic post-processing algorithms that optimize some distortion measure (e.g., PSNR, SSIM). In this paper we propose a different paradigm for JPEG artifact correction: Our method is stochastic, and the objective we target is high perceptual quality -- striving to obtain sharp, detailed and visually pleasing reconstructed images, while being consistent with the compressed input. These goals are achieved by training a stochastic conditional generator (conditioned on the compressed input), accompanied by a theoretically well-founded loss term, resulting in a sampler from the posterior distribution. Our solution offers a diverse set of plausible and fast reconstructions for a given input with perfect consistency. 
We demonstrate our scheme's unique properties and its superiority to a variety of alternative methods on the FFHQ and ImageNet datasets.

\end{abstract}

\begingroup
\newcolumntype{M}[1]{>{\centering\arraybackslash}m{#1}}
\setlength{\tabcolsep}{0pt} %
\renewcommand{\arraystretch}{0} %

\newlength{\hero}
\setlength{\hero}{0.3333\columnwidth}

\begin{figure}[t]
	\centering
    \begin{tabular}{c c c}
	    \footnotesize{Compressed (QF=5)} &
	    \footnotesize{Bahat~\etal~\cite{WhatImageExplorable2021bahat}} &
	    \footnotesize{Ours-P}\\
	   	\rule{0pt}{0.8ex}\\
	   	
	   	\centered{
	   	\begin{tikzpicture}[
	   		baseline=-2.45,
	   		spy using outlines={magnification=3, circle, height=1.5cm, width=1.5cm, yellow, every spy on node/.append style={thick}, connect spies},
	   		]
			\node[inner sep=0pt]{\adjincludegraphics[width=\hero, trim={{.065\width} 0 {.065\width} 0}, clip]{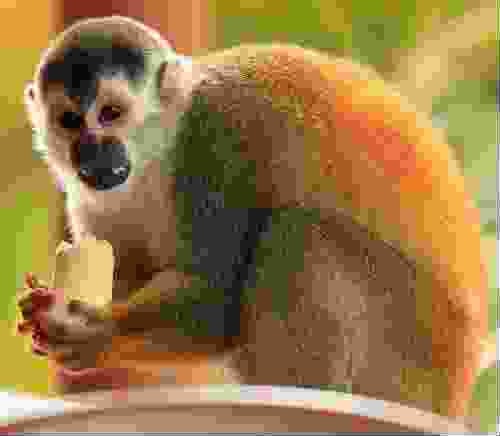}};
			\spy on (0.1,0.95) in node at (0.6,-0.6);
		\end{tikzpicture}}&
		
		\centered{
	   	\begin{tikzpicture}[
	   		baseline=-2.45,
	   		spy using outlines={magnification=3, circle, height=1.5cm, width=1.5cm, yellow, every spy on node/.append style={thick}, connect spies},
	   		]
			\node[inner sep=0pt]{\adjincludegraphics[width=\hero, trim={{.065\width} 0 {.065\width} 0}, clip]{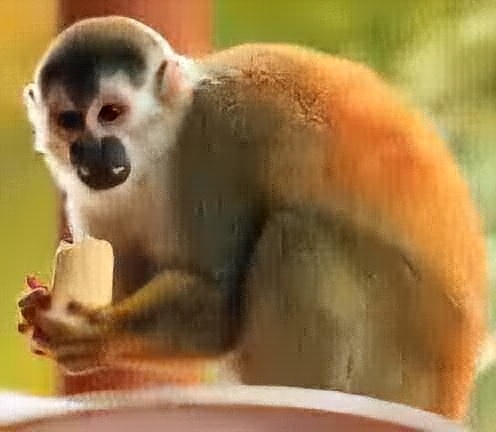}};
			\spy on (0.1,0.95) in node at (0.6,-0.55);
		\end{tikzpicture}}&

		\centered{
		\begin{tikzpicture}[
	   		baseline=-2.45,
	   		spy using outlines={magnification=3, circle, height=1.5cm, width=1.5cm, yellow, every spy on node/.append style={thick}, connect spies},
	   		]
			\node[inner sep=0pt]{\adjincludegraphics[width=\hero, trim={{.065\width} 0 {.065\width} 0}, clip]{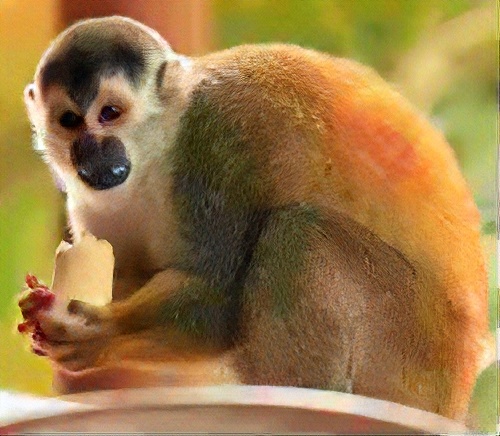}};
			\spy on (0.1,0.95) in node at (0.6,-0.55);
		\end{tikzpicture}}
		
	    \\
	    
	    \centered{
	   	\begin{tikzpicture}[
	   		baseline=-2.45,
	   		spy using outlines={magnification=2.5, circle, height=1.5cm, width=1.5cm, yellow, every spy on node/.append style={thick}, connect spies},
	   		]
			\node[inner sep=0pt]{\adjincludegraphics[width=\hero, trim={{.1\width} {.1\height} 0 {.1\height}}, clip]{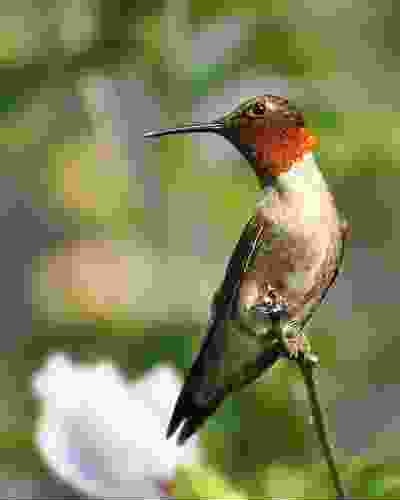}};
			\spy on (0.25,1) in node at (0.6,-0.75);
		\end{tikzpicture}}&
		
		\centered{
	   	\begin{tikzpicture}[
	   		baseline=-2.45,
	   		spy using outlines={magnification=2.5, circle, height=1.5cm, width=1.5cm, yellow, every spy on node/.append style={thick}, connect spies},
	   		]
			\node[inner sep=0pt]{\adjincludegraphics[width=\hero, trim={{.1\width} {.1\height} 0 {.1\height}}, clip]{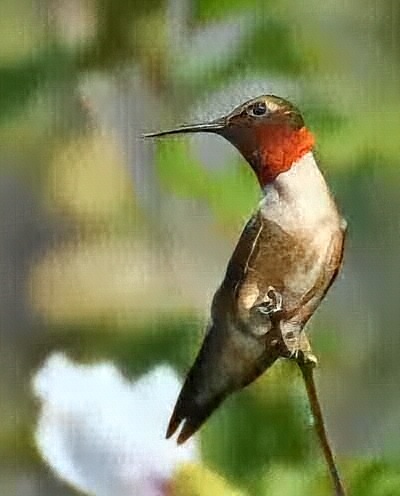}};
			\spy on (0.25,1) in node at (0.6,-0.75);
		\end{tikzpicture}}&

		\centered{
		\begin{tikzpicture}[
	   		baseline=-2.45,
	   		spy using outlines={magnification=2.5, circle, height=1.5cm, width=1.5cm, yellow, every spy on node/.append style={thick}, connect spies},
	   		]
			\node[inner sep=0pt]{\adjincludegraphics[width=\hero, trim={{.1\width} {.1\height} 0 {.1\height}}, clip]{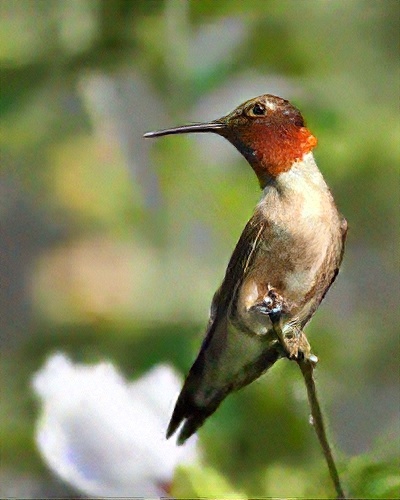}};
			\spy on (0.25,1) in node at (0.6,-0.75);
		\end{tikzpicture}}
		
	    \\
	    
	    \centered{
	   	\begin{tikzpicture}[
	   		baseline=-2.45,
	   		spy using outlines={magnification=2.5, circle, height=1.5cm, width=1.5cm, yellow, every spy on node/.append style={thick}, connect spies},
	   		]
			\node[inner sep=0pt]{\adjincludegraphics[width=\hero, trim={{.4\width} {.1\height} 0 {.1\height}}, clip]{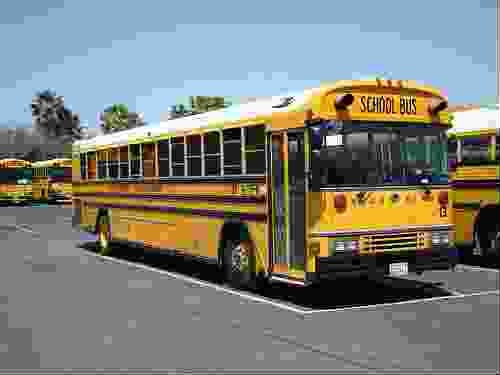}};
			\spy on (0.35,0.75) in node at (0.6,-0.6);
		\end{tikzpicture}}&
		
		\centered{
	   	\begin{tikzpicture}[
	   		baseline=-2.45,
	   		spy using outlines={magnification=2.5, circle, height=1.5cm, width=1.5cm, yellow, every spy on node/.append style={thick}, connect spies},
	   		]
			\node[inner sep=0pt]{\adjincludegraphics[width=\hero, trim={{.4\width} {.1\height} 0 {.1\height}}, clip]{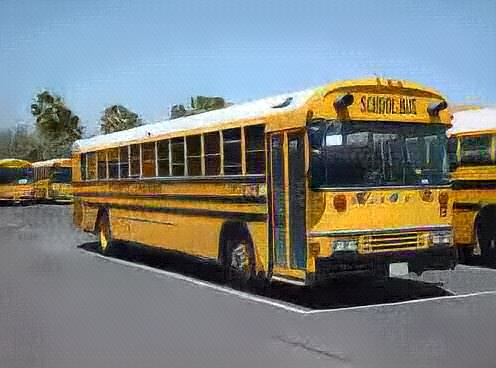}};
			\spy on (0.35,0.75) in node at (0.6,-0.6);
		\end{tikzpicture}}&

		\centered{
		\begin{tikzpicture}[
	   		baseline=-2.45,
	   		spy using outlines={magnification=2.5, circle, height=1.5cm, width=1.5cm, yellow, every spy on node/.append style={thick}, connect spies},
	   		]
			\node[inner sep=0pt]{\adjincludegraphics[width=\hero, trim={{.4\width} {.1\height} 0 {.1\height}}, clip]{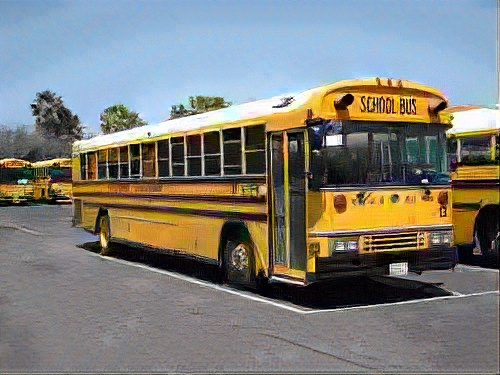}};
			\spy on (0.35,0.75) in node at (0.6,-0.6);
		\end{tikzpicture}}
		
	    \\
	    
	    \centered{
	   	\begin{tikzpicture}[
	   		baseline=-2.45,
	   		spy using outlines={magnification=1.75, circle, height=1.5cm, width=1.5cm, yellow, every spy on node/.append style={thick}, connect spies},
	   		]
			\node[inner sep=0pt]{\adjincludegraphics[width=\hero, trim={{.048\width} {.088\height} {.2\width} {.088\height}}, clip]{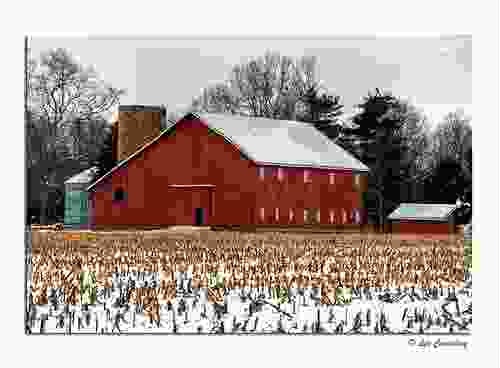}};
			\spy on (-0.9,0.65) in node at (0.6,-0.35);
		\end{tikzpicture}}&
		
		\centered{
	   	\begin{tikzpicture}[
	   		baseline=-2.45,
	   		spy using outlines={magnification=1.75, circle, height=1.5cm, width=1.5cm, yellow, every spy on node/.append style={thick}, connect spies},
	   		]
			\node[inner sep=0pt]{\adjincludegraphics[width=\hero, trim={{.048\width} {.088\height} {.2\width} {.088\height}}, clip]{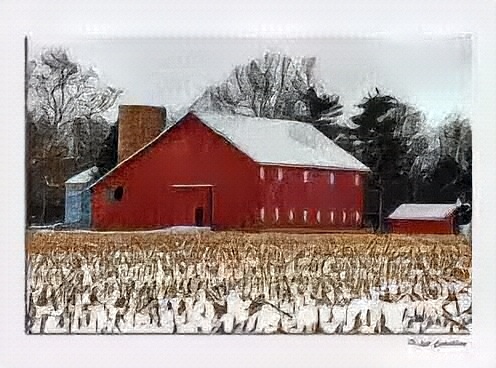}};
			\spy on (-0.9,0.65) in node at (0.6,-0.35);
		\end{tikzpicture}}&

		\centered{
		\begin{tikzpicture}[
	   		baseline=-2.45,
	   		spy using outlines={magnification=1.75, circle, height=1.5cm, width=1.5cm, yellow, every spy on node/.append style={thick}, connect spies},
	   		]
			\node[inner sep=0pt]{\adjincludegraphics[width=\hero, trim={{.048\width} {.088\height} {.2\width} {.088\height}}, clip]{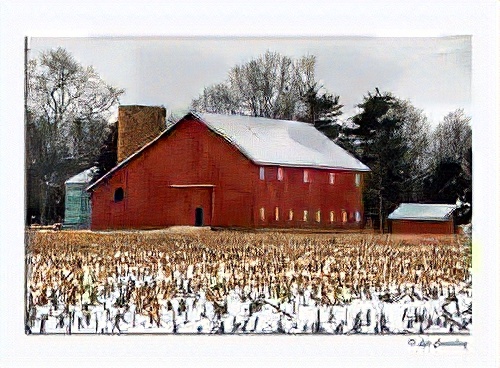}};
			\spy on (-0.9,0.65) in node at (0.6,-0.35);
		\end{tikzpicture}}
		
	    \\
	    
	    \centered{
	   	\begin{tikzpicture}[
	   		baseline=-2.45,
	   		spy using outlines={magnification=2.5, circle, height=1.5cm, width=1.5cm, yellow, every spy on node/.append style={thick}, connect spies},
	   		]
			\node[inner sep=0pt]{\adjincludegraphics[width=\hero, trim={0 0 0 0}, clip]{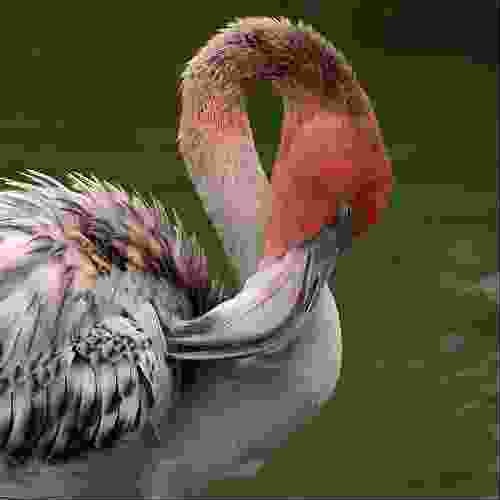}};
			\spy on (0.45,0.8) in node at (0.6,-0.6);
		\end{tikzpicture}}&
		
		\centered{
	   	\begin{tikzpicture}[
	   		baseline=-2.45,
	   		spy using outlines={magnification=2.5, circle, height=1.5cm, width=1.5cm, yellow, every spy on node/.append style={thick}, connect spies},
	   		]
			\node[inner sep=0pt]{\adjincludegraphics[width=\hero, trim={0 0 0 0}, clip]{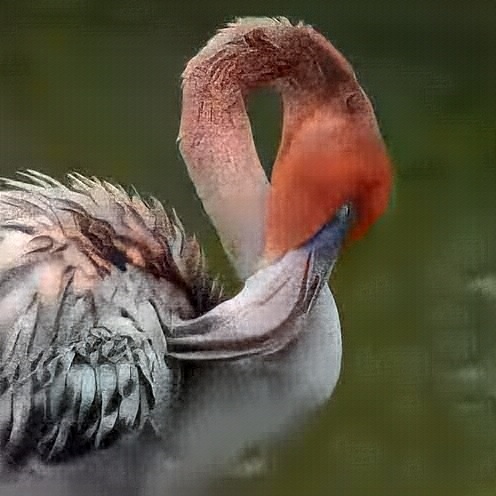}};
			\spy on (0.45,0.8) in node at (0.6,-0.6);
		\end{tikzpicture}}&

		\centered{
		\begin{tikzpicture}[
	   		baseline=-2.45,
	   		spy using outlines={magnification=2.5, circle, height=1.5cm, width=1.5cm, yellow, every spy on node/.append style={thick}, connect spies},
	   		]
			\node[inner sep=0pt]{\adjincludegraphics[width=\hero, trim={0 0 0 0}, clip]{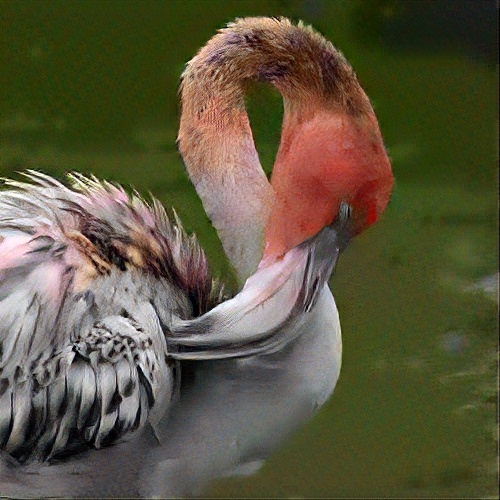}};
			\spy on (0.45,0.8) in node at (0.6,-0.6);
		\end{tikzpicture}}
		
	    \\
	    
	    \rule{0pt}{1.75ex}\\
	    
    \end{tabular}
	\caption{Reconstruction examples of highly compressed JPEG images using \cite{WhatImageExplorable2021bahat} and our proposed method.
	Our method produces stochastic outputs, all of which are perfectly consistent with the compressed image, like \cite{WhatImageExplorable2021bahat}, but with far better perceptual quality.
	}
	\label{fig:imagenet_hero}
\end{figure}
\endgroup

\section{Introduction}
\label{sec:intro}

JPEG (Joint Photographic Experts Group)~\cite{JPEGStillPicture1992wallace} is one of the most popular \emph{lossy} image compression techniques, extensively used in digital cameras, internet communications and more. JPEG reduces image file-size by discarding information that is supposed to be less valuable to a human observer. To achieve high compression ratios, JPEG often discards noticeable visual details, which may lead to strong artifacts in the decompressed image, such as blockiness.
 
Since its conception in the late 80's, numerous post-processing algorithms were proposed for removing JPEG artifacts.\footnote{While it is outside the scope of this paper to reference this vast literature, we do provide links to leading such techniques in \Cref{sec:related-work}.} Such methods start with the given compressed image and somehow provide an estimation of the source image. A common estimation approach attempts to minimize the average discrepancy between the source image and the recovered one, \eg, minimizing the Mean-Squared-Error (MSE). While this strategy may improve the quality of the compressed input and remove the blockiness effect, it still evidently leads to visually unpleasing images that are often accompanied by blurriness. As shown in~\cite{PerceptionDistortionTradeoff2018blau}, this is a direct manifestation of the perception-distortion tradeoff, which is apparently noticeable and strict for natural images and common distortion measures -- low MSE contradicts high perceptual quality.

Another key aspect in JPEG restoration processes %
is the consistency of the reconstructed image with the compressed input: The compressed version of the output image should be identical to the compressed input. This is a reasonable requirement as any inconsistent image could never be the true source. Interestingly, while older classical methods tend to be consistent (\eg. using iterative projection algorithms~\cite{ReductionCodingArtifacts1993stevenson, ImprovedImageDecompression1995orourke, ArtifactReductionLow1996luo, ImageVideoCompression1995ozcelik}), this feature has been overlooked in recent learning-based recovery techniques (such as \cite{CompressionArtifactsReduction2015dong, CASCNNDeepConvolutional2017cavigelli, BuildingDualDomainRepresentations2016guoa, QuantizationGuidedJPEG2020ehrlich}). 

In this work we propose a new JPEG reconstruction algorithm that produces consistent outputs, while also attaining very high perceptual quality. The produced images by our algorithm are free of artifacts such as blockiness, blurriness, etc. However, due to the perception-distortion tradeoff~\cite{PerceptionDistortionTradeoff2018blau}, this comes at the unavoidable cost of reduced distortion performance. Example reconstructions from our method are presented in \autoref{fig:imagenet_hero}. 
Our approach is based on a Generative Adversarial Network (GAN)~\cite{GenerativeAdversarialNets2014goodfellow}, conditioned on the compressed image, and having a stochastic synthesis process. As such, the generator is able to produce a variety of plausible and consistent reconstructions for a given compressed input. This conditional GAN is trained with a loss comprised of \emph{(i)} an adversarial term that promotes high perceptual quality, \emph{(ii)} a penalty term adopted from~\cite{HighPerceptualQuality2021ohayon} that promotes output variability (per input), and \emph{(iii)} a consistency term that promotes a compliance with the compressed input. Our optimization task theoretically admits the posterior sampler as a unique solution, leading to perfect perceptual quality and perfect consistency with the compressed input. %

The contributions of this paper are the following:
\emph{(i)} We propose a novel algorithm that produces consistent JPEG reconstructions with high perceptual quality, leading to state-of-the-art (SoTA) results on ImageNet;
\emph{(ii)} We identify an empirical perception-consistency tradeoff of stochastic estimators, extending the results presented in \cite{ReasonsSuperiorityStochastic2022ohayon}. Through a consistency penalty term (instead of enforcing perfect consistency) we are able to traverse this empirical tradeoff;
\emph{(iii)} We analyze and compare our method with previous approaches that targetted the same problem, while not being aware of this empirical tradeoff.

\section{Related work}
\label{sec:related-work}

A plethora of ideas on JPEG's artifacts removal were published since the late 80's, typically offering a post-processing mechanism within the decoding stage. These include techniques such as simple spatial filters \cite{ReductionBlockingEffect1983reeve, EnhancementJPEGCoded1995kundu, AdaptivePostfilteringTransform2001chen}, MSE optimization of codebooks \cite{PracticalRealtimePostprocessing1996hong, ImprovedDecoderTransform1992wu}, MAP optimization using Gibbs priors \cite{ReductionCodingArtifacts1993stevenson, ImageVideoCompression1995ozcelik, CodingArtifactRemoval1997li}, sparse representations \cite{ImageDeblockingSparse2012jung, ReducingArtifactsJPEG2014chang} and many other techniques (\eg, see the review by Shen and Kuo~\cite{ReviewPostprocessingTechniques1998shen}).

In recent years, deep-learning-based methods took the lead in post-processing JPEG compressed images. Dong \etal~\cite{CompressionArtifactsReduction2015dong} were the first to suggest a CNN-based regression model, followed by work such as \cite{CASCNNDeepConvolutional2017cavigelli} that improved the architecture and the loss function. Guo \etal~\cite{BuildingDualDomainRepresentations2016guoa} combined pixel domain and DCT domain sub-networks to take advantage of JPEG's mode of operation, followed by work such as \cite{D3DeepDualDomain2016wang, DmcnnDualDomainMultiScale2018zhang} that improved the dual-domain concept and \cite{ReductionJPEGCompression2020sun, QuantizationGuidedJPEG2020ehrlich} that took a step further by operating solely in the DCT domain. All of these methods adopt a supervised approach, training a network to best fit a given compressed image to its desired ideal output. 

As noted in \Cref{sec:intro}, a common difficulty of these techniques (classical or deep-learning based) is the overly smoothed reconstructed images that lack fine texture and high frequency details, which occurs due to the inherit tradeoff between perceptual quality and an optimization of a distortion criterion \cite{PerceptionDistortionTradeoff2018blau}. Indeed, some methods \cite{OneToManyNetworkVisually2017guo, DeepUniversalGenerative2019galteri, QuantizationGuidedJPEG2020ehrlich} incorporated adversarial training \cite{GenerativeAdversarialNets2014goodfellow} to produce sharp details while not being aware of this tradeoff, hence kept incorporating distortion measures as part of their loss, and thus compromising on the perceptual quality of their outcome as well.

Among the deep-learning-based works, only a few addressed the desire that the reconstructed images should be consistent with the compressed inputs. The work reported in \cite{ReductionJPEGCompression2020sun} enforced this requirement as a constraint on the resulting DCT coefficients, while \cite{OneToManyNetworkVisually2017guo} encouraged the reconstructions to obey this requirement using a penalty as part of the optimization criterion.

We now turn to discuss two recent and relevant works that inspired this project ~\cite{HighPerceptualQuality2021ohayon, WhatImageExplorable2021bahat}. PSCGAN~\cite{HighPerceptualQuality2021ohayon} is a recently proposed image denoiser that attempts to attain a stochastic estimator that samples from the posterior distribution. While \cite{HighPerceptualQuality2021ohayon} bares some similarities to our work, it differs in the task addressed (denoising vs. JPEG reconstruction), the conditional GAN loss used \cite{GenerativeAdversarialNets2014goodfellow, WhichTrainingMethods2018mescheder}, the fact that it disregards consistency, and the very different neural network architectures deployed. Nevertheless, \cite{HighPerceptualQuality2021ohayon} introduce a first moment loss that encourages acceptable distortion but bypasses the perception-distortion tradeoff, which we leverage in our JPEG recovery algorithm. 

The work by Bahat \etal~\cite{WhatImageExplorable2021bahat} deserves a special mention as it is the only prior work, to the best of our knowledge, that addresses both the perception-distortion tradeoff and the consistency with the compressed input. \cite{WhatImageExplorable2021bahat} imposes consistency by predicting bounded residuals in the DCT domain and optimizing a GAN loss that bypasses a distortion criteria, hence avoiding the pitfall of the mentioned tradeoff. 
Nonetheless, our proposed approach achieves much better perceptual quality results compared to \cite{WhatImageExplorable2021bahat} while retaining perfect consistency. We give an explanation to this improvement in \Cref{sec:practice} based on the empirical perception-consistency tradeoff, and conduct extensive experiments to compare both methods in \Cref{sec:experiments}.

\section{Method: fundamentals}
\label{sec:fundementals}

We assume that a natural image $X$ is a multivariate random variable with probability density function $p_{X}$. We denote by $Y$ the JPEG-compressed-decompressed version of $X$, and from $Y$ we provide an estimate of $X$, denoted by $\hat{X}$.
We do so by attempting to sample from the posterior distribution $p_{X|Y}$.
Such an estimator would be highly effective since 
\emph{(i)} it's outputs are consistent with the measurements\footnote{In our notations throughout the paper, $Y=\jpeg(X)$ implies a compression-decompression operation, and thus $Y$ is an image of the same size as $X$.}, i.e. $\jpeg(\hat{X})=Y$; and
\emph{(ii)} it attains perfect perceptual quality\cite{PerceptionDistortionTradeoff2018blau}, i.e., $p_{\hat{X}}=p_X$.

\subsection{JPEG}

The JPEG compression algorithm \cite{JPEGStillPicture1992wallace} operates using a pre-defined quantization matrices $Q \in \mathbb{Z}^{8 \times 8}$ in the DCT domain on an image $X$ in the YCbCr color-space, which, for simplicity, we assume is of size $8m \times 8n \times 3$.
The algorithm starts by dividing the image's channels into $8\times 8$ blocks $\{X_i\}^{m \times n \times 3}$. Each block $X_i$ is transformed by the 2D-DCT transformation $X_i^D=\dct(X_i)$ and divided elementwise by the corresponding quantization matrix $Q$ to achieve $X_i^Q=X_i^D \oslash Q$. Finally, each entry of the block is rounded to achieve $X_i^R=\round{X_i^Q}$. All the blocks $\{X_i^R\}^{m \times n \times 3}$ are stored using a lossless entropy-coding algorithm alongside the matrix $Q$.
Decompression is obtained by multiplying these rounded values by $Q$ and applying an inverse 2D-DCT on the resulting blocks.
We denote the above compression-decompression process by $\jpeg_Q(\cdot)$.

Note that rounding is a lossy operation, and the matrices $Q$ controls the amount of lost information. These matrices are a function of the quality-factor (QF) chosen by the user -- an integer between 0 to 100, where 0 means maximum compression (in this paper we use the baseline matrices defined in \cite{Recommendation81Information1992}).

\subsection{Consistency}

To measure the consistency of $\hat{X}$ with $Y$, we define the consistency index by
\begin{equation}
\label{eq:consistency_index}
    c(p_{\hat{X}}, p_Y) = \mathbb{E}_{Y} \left[\mathbb{E}_{\hat{X}|Y} \left[\norm{Y - \jpeg_Q(\hat{X})} | Y \right]\right],
\end{equation}
The estimator $\hat{X}$ is perfectly consistent with the compressed inputs $Y$ if $Y=\jpeg_Q(\hat{X})$, or equivalently, iff $c(p_{\hat{X}}, p_Y)=0$.
Interestingly, as we show next, prior JPEG restoration models that attempt to minimize the MSE loss implicitly striven (at least theoretically) to become perfectly consistent, since the MMSE estimator produces consistent restoration:

\begin{thmE}[][end, restate, no link to theorem, text link={See proof in \pratendSectionlikeCref.}]
\label{thm:mmse}
Let $\bar{X}(Y)$ be an MMSE estimator for JPEG artifact removal, i.e $\bar{X}(Y)=\mathbb{E} \left[ X | Y \right]$. Then $\bar{X}(Y)$ is necessarily perfectly consistent with the compressed input $Y$.
\end{thmE}
\begin{proofE}
Denote by $D$ the matrix that performs block-wise $\dct$ and elementwise division by the matrix $Q$. Then $\bar{X}(Y)$ is consistent with $Y$ iff $ \norm{D \bar{X}(Y) - D Y}_\infty \leq \frac{1}{2} $.
And indeed,
\begin{equation*}
    \begin{aligned}
	    \norm{D \bar{X}(Y) - D Y}_\infty &= \norm{D \mathbb{E} \left[ X | Y \right] - D Y }_\infty \\
	    &= \norm{\mathbb{E} \left[ DX - DY | Y \right] }_\infty \\
	    &\leq \mathbb{E} \left[ \norm{ DX - DY }_\infty | Y \right] \\
	    &\leq \mathbb{E} \left[ \frac{1}{2} | Y \right] = \frac{1}{2},
    \end{aligned}
\end{equation*}
where we used the triangle inequality and the fact that at any point where $p(X|Y) > 0$, the maximal difference in the DCT domain, before rounding, is $\frac{1}{2}$.
\end{proofE}

\subsection{Perceptual quality}
While there are several definitions for the notion of perceptual quality, we follow the notion developed in~\cite{PerceptionDistortionTradeoff2018blau}, which measures for an estimator $\hat X$ the index 
\begin{equation}
\label{eq:perceptual_index}
    d(p_X, p_{\hat{X}}),
\end{equation}
where $d(p,q)$ is some divergence between the two distributions, \eg, Kullback-Leibler's (KL), Jensen-Shanon's (JS) divergence, or Wasserstein distance. 
$\hat{X}$ attains perfect perceptual quality when $d(p_X, p_{\hat{X}})=0$, i.e., $p_X = p_{\hat{X}}$.

\subsection{Posterior sampling - our goal}

The work in \cite{ReasonsSuperiorityStochastic2022ohayon} ties the perceptual quality and the consistency of estimators via the following theorem:

\begin{thm}\label{thm:posterior-optimal}
Let $X \sim p_X$ be a random multivariate variable, and let $Y=D(X)$ be a deterministic degradation of $X$.
If $\hat{X}$ is an estimator such that $p_{X}=p_{\hat{X}}$ and $Y=D(\hat{X})$, then $p_{\hat{X}|Y}=p_{X|Y}$, i.e., $\hat{X}$ is a necessarily posterior sampler.
\end{thm}
\begin{proof}
Found in \cite{ReasonsSuperiorityStochastic2022ohayon}.
\end{proof}
In our case, $Y=D(X)=\jpeg_Q(X)$.
Recall that $X$ cannot be uniquely recovered from $Y$ since $\jpeg_Q(\cdot)$ is a non-invertible degradation, and thus the support of $p_{X|Y}$ is not a singleton -- there is a variety of possible sources that correspond to the same compressed image. Since we aim to sample from the posterior, our method must be stochastic, capable of providing many reconstructed samples given the same compressed image it. This is in opposition to most prior work, which adopts a deterministic recovery strategy. 

Our solution is comprised of a conditional GAN~\cite{ConditionalGenerativeAdversarial2014mirza} that takes $Y$ as an input and produces high perceptual quality outputs that are consistent with $Y$. Posing our task as finding a sampler from $p_{\hat{X}|Y}$, we form a loss function that has two ingredients. Just as with all GAN methods, we use an adversarial loss $\mathcal{L}_\text{Adv}$ to minimize the divergence between the real and the generated data, matching $p_X$ and $p_{\hat{X}}$ as best as possible.
To promote consistency, we include another objective that penalizes any discrepancy between $Y$ and $\jpeg_Q(\hat{X})$, leading to the following final loss:
\begin{equation}
\label{eq:uncons_opt}
\begin{gathered}
    \min_{p_{\hat{X}|Y}} \mathcal{L}_\text{Adv}(p_{X}, p_{\hat{X}}) \\
    + \lambda \mathbb{E}_{Y} \mathbb{E}_{\hat{X}|Y} \left[ \norm{Y - \jpeg_Q(\hat{X})}_2^2 \right].
\end{gathered}
\end{equation}
According to \autoref{thm:posterior-optimal}, \autoref{eq:uncons_opt} admits a single optimal solution for any $\lambda>0$: $p_{X|Y}$. This is true since any optimal solution would attain perfect perceptual quality and produce perfectly consistent restorations.
Practically, however, we solve \autoref{eq:uncons_opt} with parametric neural networks and with a data set of finite size, and thus our solution may not be the true posterior. Moreover, due to the nature of practical optimization, the choice of $\lambda$ may also affect the obtained solution, as discussed in the following section.

\section{Method: practice}
\label{sec:practice}

\subsection{Achieving the posterior}
\label{sec:practice.1}

The work in \cite{ReasonsSuperiorityStochastic2022ohayon} revealed that a posterior sampler is the only consistent restoration algorithm that attains perfect perceptual quality.
This leads to a tradeoff between consistency and perceptual quality for deterministic estimators, as any such estimator cannot be a posterior sampler.
This theoretical tradeoff \textbf{does not} affect our method, as our algorithm is stochastic.
 In practice, however, it is very likely that a suboptimal optimization procedure, a highly non-convex loss surface, a finite size data set, and a limited capacity architecture, would all make it extremely hard to attain the perfect solution -- a sampler from the posterior. Thus, we expect to improve both the consistency index~(\autoref{eq:consistency_index}) and the perceptual index~(\autoref{eq:perceptual_index}) up to a certain point, beyond which we shall observe an \emph{empirical} tradeoff, where the improvement of one quality comes at the expense of the other.
 This empirical tradeoff is not revealed or discussed in \cite{ReasonsSuperiorityStochastic2022ohayon}.
 By changing $\lambda$ in our optimization task (\autoref{eq:uncons_opt}), such a tradeoff can be controlled, i.e., we can decide to attain higher perceptual quality or better consistency.

As such an empirical tradeoff has not been revealed in prior work on stochastic estimators, the balance between perceptual quality and consistency has been implicitly addressed. Bahat \etal~\cite{WhatImageExplorable2021bahat} imposed a consistency requirement as a constraint using their generator architecture and not as a penalty. This can be interpreted as choosing $\lambda\rightarrow\infty$, requiring perfect consistency at the cost of lower perceptual quality. On the other hand, prior work that attempted to attain the posterior without paying attention to consistency in any way, such as \cite{HighPerceptualQuality2021ohayon}, controlled the tradeoff implicitly by the choice of architecture, loss, and optimizers.
In fact, we argue that any attempt to sample from the posterior would most likely compromise on either consistency or perceptual quality, since attaining the true posterior is highly challenging.

\subsection{Training method}
\label{sec:practice.train}

We denote our estimator by $G_\theta(Z,Y)$, where $G_{\theta}$ is a neural network, $Y$ is the input JPEG-compressed image, and $Z$ is a random seed that enables a diverse set of outputs for any input image $Y$.
Our training procedure consists of a weighted sum of several objectives.
First, a non-saturating adversarial loss term~\cite{GenerativeAdversarialNets2014goodfellow}, accompanied by an  $R_{1}$ gradient penalty for the critic~\cite{WhichTrainingMethods2018mescheder}.
We denote these GAN losses by $V(D_{\omega},G_{\theta})$ and $R_{1}(D_{\omega})$, where $D_{\omega}$ is our critic.
Second, we use a consistency penalty term $C(G_\theta)$, as in \autoref{eq:uncons_opt}:
\begin{equation}
    \begin{aligned}
        C(G_\theta) =& \mathbb{E}_{Y,Z} \left[ \norm{Y - \jpeg_Q(G_\theta(Z, Y))}_2^2 \right].
    \end{aligned}
\end{equation}
Third, we incorporate a first-moment penalty term $FM(G_\theta)$, originally proposed in \cite{HighPerceptualQuality2021ohayon}:
\begin{equation}\label{eq:first-moment}
    \begin{aligned}
        FM(G_\theta) =& \mathbb{E}_{X,Y} \left[ \norm{X-\mathbb{E}_{Z}[G_\theta(Z, Y)|Y]}_2^2 \right],
    \end{aligned}
\end{equation}
which specifies that the average of many outputs $G_\theta(Z,Y)$ that refer to a fixed $Y$ while varying $Z$ should be close to the ideal image $X$. If indeed this multitude of outputs form a fair sampling from the posterior, this penalty leads exactly to the MMSE estimation -- 
$ \mathbb{E}_Z \left[ G_\theta (Z,Y) |Y\right] = \mathbb{E}_X \left[ X | Y\right]$.
This term is a natural force that replaces the more intuitive supervised distortion penalty $\mathbb{E}_{X,Z} \left[ \norm{X - G_\theta(Z,Y)}_2^2 \right]$. As we have already mentioned, a distortion penalty typically hinders the perceptual quality, while  the alternative in \autoref{eq:first-moment} does not, further strengthening the overall optimization.
More specifically, without using \autoref{eq:first-moment} we observe mode-collapse during training -- as we only have one $X$ per given $Y$ in our dataset, the generator is not incentivized to generate stochastic reconstructions, hence, it almost completely ignores $Z$ (as explained in~\cite{HighPerceptualQuality2021ohayon}).

On top of the above, in complex general content datasets such as ImageNet~\cite{ImageNetLargescaleHierarchical2009deng} we empirically find that further guidance is required in order to achieve satisfactory results.
In these scenarios we include a VGG ``perceptual'' loss \cite{VeryDeepConvolutional2015simonyan, PerceptualLossesRealTime2016johnson}, which promotes the generation of fine details in severely compressed images:
\begin{equation}
    \begin{aligned}
        P(G_\theta) =& \mathbb{E}_{X,Y,Z} \left[ \norm{\vgg(X)-\vgg(G_\theta(Z, Y))}_2^2 \right].
    \end{aligned}
\end{equation}
Where $\vgg(\cdot)$ are the features of a trained VGG-19 network at the specified convolutional layer. Lastly, in order to increase the variation in the estimator's reconstructions, we introduce a new second-moment penalty:
\begin{equation}
    \begin{aligned}
        SM&(G_\theta) = \mathbb{E}_{X,Y} \left[ \norm{ \left(X-\bar{X}(Y)\right)^2 - \var_{Z}[G_\theta(Z, Y)|Y]} \right].
    \end{aligned}
\end{equation}
This term specifies that the variance of the generated images for a given $Y$ should be close to the sample variance using a single {\bf ground-truth} sample and a pre-trained MMSE estimator $\bar{X}(Y)$. In \Cref{sec:second_moment} we give further rational behind this penalty.

In \Cref{sec:ablation} we present an ablation study to show the importance of the VGG loss and the second-moment penalty. Note that while we do not have a theoretical guaranty that the VGG loss does not introduce a perception-distortion tradeoff, we see in our experiments that both the perceptual quality and the consistency of the generated images improves.

All of the above forces result in the following unified minimax game:
\begin{equation}
    \label{eq:opt}
    \begin{aligned}
	    \min_\theta \max_\omega  & V(D_\omega, G_\theta) + \lambda_{R_1} R_1(D_\omega) + \lambda_{C} C(G_\theta) \\ & + \lambda_{FM} FM(G_\theta) + \lambda_{P} P(G_\theta) + \lambda_{SM} SM(G_\theta).
    \end{aligned}
\end{equation}
We solve this task using a block-coordinate optimization, resulting in alternating optimization tasks:
\begin{equation}
    \label{eq:opt_d}
    \begin{aligned}
        \max_\omega V(D_\omega, G_\theta) + \lambda_{R_1} R_1(D_\omega),
    \end{aligned}
\end{equation}
\begin{equation}
    \label{eq:opt_g}
    \begin{aligned}
        \min_\theta & V(D_\omega, G_\theta) + \lambda_{C} C(G_\theta) + \lambda_{FM} FM(G_\theta) \\ & + \lambda_{P} P(G_\theta) + \lambda_{SM} SM(G_\theta).
    \end{aligned}
\end{equation}
We should note that the consistency penalty term requires a differentiable implementation of JPEG, a concept introduced in prior work such as \cite{BetterCompressionDeep2021talebi, RateDistortionAccuracyTradeoffJPEG2020luo, shin2017jpeg}. We opt to approximate the backward pass of the rounding operation using $\nabla \round{X} = 1$.

\subsection{Projection}
\label{sec:practice-proj}

As we enforce consistency through the training objective and not via the architecture, we can expect the results of our trained models to be inconsistent to some degree. 
Post training, though, we can still produce perfect consistency by projecting the DCT coefficients of any reconstructed image $\hat{X}$ to the range permitted by the rounding operation that created $Y$.
We denote the projected results by $\tilde{X}$. Per block, and in the DCT domain, the projection operation is defined as
\begin{equation}
  \tilde{X}_i^Q = Y_i^Q + \max\left( \min\left(\hat{X}_i^Q-Y_i^Q, 0.5 \right), -0.5 \right).
\end{equation}
Note that it is expected that our perceptual quality would degrade as a result of the projection operation, since it is not guaranteed to result in a natural image and also due to empirical perception-consistency tradeoff for stochastic estimators.
However, for a model that attains high perceptual quality and satisfying consistency, this degradation should be minor. We demonstrate the projection's effect on our models in \Cref{sec:experiments}.

\section{Experiments}
\label{sec:experiments}

\noindent {\bf Methods:}
We compare our method (denoted \textbf{Ours}) with several alternative post-processing methods:
\emph{(1)} \textbf{QGAC} and \textbf{QGAC-GAN}~\cite{QuantizationGuidedJPEG2020ehrlich}, a SoTA regression method and a GAN method fine-tuned from it;
\emph{(2)} \textbf{SwinIR}~\cite{SwinIRImageRestoration2021lianga}, a SoTA regression methods trained separately for each QF;
\emph{(3)} \textbf{FBCNN}~\cite{FlexibleBlindJPEG2021jiang}, a SoTA regression method;
\emph{(4)} \textbf{Bahat \etal}~\cite{WhatImageExplorable2021bahat}, as noted in \Cref{sec:related-work}, the only other work that attempts to attain perfect consistency and perceptual quality;
and \emph{(5)} \textbf{Ours\=/MSE}, our very same architecture trained as a regression model using solely an MSE loss as a baseline. Following \autoref{thm:mmse}, this model should produce consistent reconstructions.
Moreover, we test the results obtained by our method after projection (as explained in \Cref{sec:practice-proj}), and denote these by \textbf{Ours\=/P}.

Unless mentioned otherwise, when possible we use checkpoints as published by the authors. In other cases that require training, we use commonly available hardware.\footnote{The networks were trained on a single NVIDIA A6000 GPU.}
Please refer to \Cref{sec:imp} for details on training and architectures.

\noindent {\bf Metrics:}
To measure consistency, we compute the RMSE between the compressed-decompressed versions of the original and the restored images using the same JPEG settings. Note that the value shown is per-pixel and in units of gray-levels.

To evaluate perceptual quality we compute the FID~\cite{GANsTrainedTwo2017heusel, Seitzer2020FID} between the real uncompressed images and the restored ones.
For stochastic methods (ours and Bahat~\etal's) we present the mean and standard deviation of 64 repeated FID evaluations, where in each we generate one restoration per compressed test image.
This makes sure that our model consistently performs with high perceptual quality regardless of the seed.
We also present a ``ground truth'' score, which is the FID between the training and the validation images (in \Cref{sec:experiments-results-ffhq}) or between the validation and the test images (in \Cref{sec:experiments-results-general}).

\subsection{Results: FFHQ}
\label{sec:experiments-results-ffhq}

\colorlet{myYellow}{yellow!50!white}
\colorlet{myPink}{pink!50!white}
\colorlet{myBlue}{cyan!50!white}
\colorlet{myOrange}{orange!50!white}
\colorlet{myRed}{red!50!white}
\colorlet{myGrey}{black!30!white}

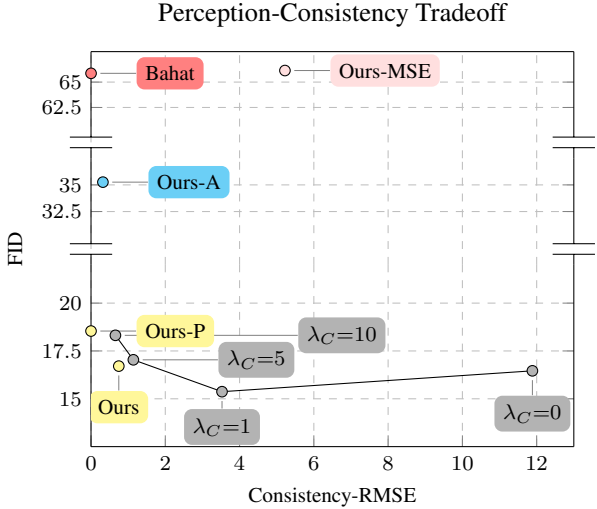
\begin{figure}[t]
\centering
\begin{tikzpicture}
\tikzset{
    every pin/.style={fill=myYellow,rectangle,rounded corners=3pt},
  }
\begin{scope}[local bounding box=graph]
\begin{groupplot}[
    group style={
        group size=1 by 3,
        horizontal sep=0pt,
        vertical sep=0pt,
    },
    width=8cm,
    height=5.5cm,
    ymin=10, ymax=25,
    xmin=0, xmax=15,
    ytick={0, 1, 3.5, ..., 13},
    xmajorgrids=true, ymajorgrids=true,
    log ticks with fixed point,
    grid style=dashed,
]

\pgfplotsset{every tick label/.append style={font=\footnotesize}}

\nextgroupplot[
	ymin=57.5, ymax=68,
	xmin=0, xmax=13,
	xtick style={draw=none},
	axis y line=box,
    axis x line*=top,
    height=3cm,
	axis y discontinuity=parallel,
	ytick={62.5, 65},
	xmajorticks=false,
]

\addplot [only marks, fill=myPink, draw=black] coordinates {
    (5.2217, 66.13762678661595)
} node[pin={[fill=myPink]0:\footnotesize Ours\=/MSE}]{};

\addplot [only marks, fill=myRed, draw=black] coordinates {
    (0.001, 65.86222994431631)
} node[pin={[fill=myRed]0:\footnotesize Bahat}]{}; 

\nextgroupplot[
	ymin=27.5, ymax=37.5,
	xmin=0, xmax=13,
	separate axis lines,
	axis y line=box,
    x axis line style={draw opacity=0},
    height=3cm,
	ytick={32.5, 35},
	xmajorticks=false,
	axis y discontinuity=parallel,
]

\addplot [only marks, fill=myBlue, draw=black] coordinates {
    (0.3203, 35.25803278273065)
} node[pin={[fill=myBlue]0:\footnotesize Ours-A}]{};                

\nextgroupplot[
	ymin=12.5, ymax=22,
	xmin=0, xmax=13,
	axis x line*=bottom,
    axis y line=box,
    ytick={15, 17.5, ..., 20},
    height=4cm,
    xlabel={\footnotesize Consistency-RMSE},
]

\addplot [mark=none] coordinates {
	(11.8848, 16.460969247391553)
	(3.5275, 15.371075465842333)
    (1.1402, 17.036795687266892)
    (0.6573, 18.316255190337376)
};

\addplot [only marks, align=center, fill=myGrey, draw=black] coordinates {
    (11.8848, 16.460969247391553)
} node[pin={[fill=myGrey, pin distance=0.2cm]270:\footnotesize $\lambda_C{=}0$}]{};

\addplot [only marks, fill=myGrey, draw=black] coordinates {
    (3.5275, 15.371075465842333)
} node[pin={[fill=myGrey, pin distance=0.1cm]270:\footnotesize $\lambda_C{=}1$}]{};

\addplot [only marks, fill=myGrey, draw=black] coordinates {
    (1.1402, 17.036795687266892)
} node[pin={[fill=myGrey, pin distance=1cm]0:\footnotesize $\lambda_C{=}5$}]{};

\addplot [only marks, fill=myGrey, draw=black] coordinates {
    (0.6573, 18.316255190337376)
} node[pin={[fill=myGrey, pin distance=2.3cm]0:\footnotesize $\lambda_C{=}10$}]{};

\addplot [only marks, fill=myYellow, draw=black] coordinates {
    (0.7481, 16.707357759891817)
} node[pin={[fill=myYellow, pin distance=0.2cm]270:\footnotesize Ours}]{};

\addplot [only marks, fill=myYellow, draw=black] coordinates {
    (0, 18.540371937003314)
} node[pin={[fill=myYellow]0:\footnotesize Ours-P}]{};       

\end{groupplot}
\end{scope}
\node (title) at ($(group c1r1.north)!0.5!(group c1r1.north)$) [above, yshift=\pgfkeysvalueof{/pgfplots/every axis title shift}] {Perception-Consistency Tradeoff};
\node (ylabel) at ($(group c1r1.north west)!0.5!(group c1r3.south west)$) [left=1cm, xshift=\pgfkeysvalueof{/pgfplots/every axis title shift}] {\begin{turn}{90}\footnotesize{FID}\end{turn}};

\end{tikzpicture}
\caption{FID versus Consistency of different methods on FFHQ-128 with QF=5. By adjusting $\lambda_C$, our method exposes explicit control over the perception-consistency tradeoff of the trained estimator. Observe that when $\lambda_C$ is increased, the consistency of the estimators improves up to a point where consistency comes at the expense of perceptual quality. By carefully tuning $\lambda_C$ during training we are able to reach an estimator (Ours) with similar perceptual quality as the quality attained when using $\lambda_C{=}0$, but with a much improved consistency. The consistency of constrained methods (Ours-P, Bahat~\cite{WhatImageExplorable2021bahat}) are practically zero -- see \Cref{sec:experiments-results-ffhq} for more details.}
\label{fig:ffhq_methods}
\end{figure}

To showcase the empirical perception-consistency tradeoff we use the FFHQ \cite{StyleBasedGeneratorArchitecture2019karras} thumbnails dataset, in which each image is of size $128\times128$.
 We compress the images with QF=5 and use the same train-validation-test split as in \cite{HighPerceptualQuality2021ohayon}. We compare our method to Bahat~\etal (trained using their official implementation and following the training method described in their paper) and the baseline Ours\=/MSE method.
	To show the importance of the consistency regularization, we also compare the performance of our method trained by setting $\lambda_C=0$ in the loss formulation (\autoref{eq:opt_g}).
	As a proxy to an MMSE estimator, we also compare the performance of an average of 64 realizations per input from our model, denoted as \textbf{Ours\=/A}.
The empirical perception-consistency tradeoff of the different methods on the FFHQ-128 dataset is visualized in \autoref{fig:ffhq_methods}.
Further visual results and quantitative FID, consistency and PSNR results are summarized in \Cref{sec:more_results} (\autoref{tab:ffhq}).

\noindent {\bf Consistency:}
As shown in \autoref{fig:ffhq_methods}, Ours with $\lambda_C{=}0$ and Ours\=/MSE both produce unsatisfactory consistency levels, which suggests that to attain high consistency, some type of supervision is required. Just by activating our penalty term we are able to improve the consistency-RMSE by more than 11 gray-levels.

Note that both Ours\=/A and Ours\=/MSE are approximations of an MMSE estimator, and they differ only in their loss functions. 
While Ours\=/MSE attains slightly higher PSNR (see \Cref{sec:more_results}), Ours\=/A attains significantly better consistency.
Following \autoref{thm:mmse} we know that an MMSE estimator should be perfectly consistent, so Ours\=/A is closer to a true MMSE estimator in that sense, even thought it attains a slightly lower PSNR compared to the regression model.

By projecting the restored images (Ours-P) we improve the consistency significantly, bringing our method to be on-par with Bahat's. While the projection should have resulted in perfect-consistency, we get a slight deviation due to numerical approximations in the JPEG algorithm (in the color-space conversion). This is also apparent in the results of Bahat's method that enforce the consistency as part of the architecture. In \Cref{sec:numerical-erros} we further investigate this phenomenon and show that it affects even the standard JPEG implementation \emph{libjpeg}~\cite{IndependentJPEGGroup}.

In \Cref{sec:more_results} we present more visual and quantitative results regarding the consistency of the different methods.

\noindent {\bf Perception-consistency tradeoff:}
By controlling $\lambda_C$ in \autoref{eq:opt_g} we can incentivize our generator to produce more consistent reconstructions with the compressed input, as evident in \autoref{fig:ffhq_methods}.
Starting with $\lambda_C{=}0$, we converge to an estimator with good perceptual quality but with lacking consistency. %
By choosing a small penalty, such as $\lambda_C{=}1$, we are able to converge to an estimator that achieves both better perceptual quality and better consistency. Yet, the consistency of the results are far from being satisfactory.
Cranking the penalty coefficient up to $\lambda_C{=}5$ improves the consistency significantly, but the perceptual-quality deteriorate compared to $\lambda_C{=}0$. This trend continues as we further increase $\lambda_C$.

By carefully adjusting $\lambda_C$ during training (please refer to \Cref{sec:imp-train} for details) we are able to achieve a balanced result -- the perceptual quality of $\lambda_C{=}0$ with the improved consistency of a large penalty term.
This exploration demonstrates our main point in \Cref{sec:practice.1} -- the chosen penalty term gives us explicit control over the empirical perception-consistency tradeoff, which lets us converge to a better local-minima, as opposed to the implicit or no control in previous methods. We can expect further improvements with more hyper-parameter tuning.

\subsection{Results: general content}
\label{sec:experiments-results-general}

To showcase the performance of our method we train our model on $128\times128$ patches extracted from DIV2k~\cite{NTIRE2017Challenge2017agustsson} and Flickr2k~\cite{NTIRE2017Challenge2017timofte} datasets at multiple QFs in the range $[5, 50]$ and test all the mentioned methods on the ImageNet-ctest10k dataset, as proposed in \cite{PaletteImagetoImageDiffusion2022sahariaa}.
QGAC~\cite{QuantizationGuidedJPEG2020ehrlich} and FBCNN~\cite{FlexibleBlindJPEG2021jiang} are also trained on DIV2K and Flickr2K, SwinIR is also trained on BSDS500~\cite{ContourDetectionHierarchical2011arbelaez} and WED~\cite{WaterlooExplorationDatabase2017ma}, while Bahat~\etal~\cite{WhatImageExplorable2021bahat} is trained on ImageNet~\cite{ImageNetLargescaleHierarchical2009deng}.
In \autoref{fig:imagenet_methods} we present the FID, consistency and PSNR of the different methods across a range of QFs from 5 to 50 on ImageNet-ctest10k, and in \Cref{fig:imagenet_hero,fig:imagenet_compare} we present reconstruction examples from ImageNet-ctest10k of some of the different methods.
Further visual results on ImageNet, LIVE1 \cite{LIVEImageQualitysheikh, StatisticalEvaluationRecent2006sheikh} and BSDS500 are presented in \Cref{sec:more_results}.

\colorlet{myYellow}{yellow!50!white}
\colorlet{myGreen}{green!50!white}
\colorlet{myPink}{pink!50!white}
\colorlet{myOrange}{orange!50!white}
\colorlet{myRed}{red!50!white}
\colorlet{myGrey}{black!30!white}
\definecolor{mypink2}{RGB}{219, 48, 122}

\definecolor{myblue1}{HTML}{a6cee3}
\definecolor{myblue2}{HTML}{1f78b4}
\definecolor{mygreen1}{HTML}{b2df8a}
\definecolor{mygreen2}{HTML}{33a02c}
\definecolor{myred1}{HTML}{ff7f00}
\definecolor{myred2}{HTML}{e31a1c}
\definecolor{myorange}{HTML}{fdbf6f}
\definecolor{myyellow}{HTML}{f2bd9e}

\begin{figure*}[t]
\centering
\begin{tikzpicture}
\tikzset{
    every pin/.style={fill=myYellow,rectangle,rounded corners=3pt},
  }
\begin{scope}[local bounding box=graph]
\begin{groupplot}[
	group style={
        group size=3 by 1,
        horizontal sep=1.35cm,
    },
    width=6.2cm,
    height=5cm,
    xmin=0, xmax=55,
    xlabel={\footnotesize QF},
    xtick={5, 10, 20, ..., 50},
    xmajorgrids=true, ymajorgrids=true,
    log ticks with fixed point,
    grid style=dashed,
    legend style ={
        draw=white, 
        fill=white,
        legend columns=4,
        at={(0.5, -0.25)}, anchor=north
    },
    legend cell align={left},
]

\pgfplotsset{every tick label/.append style={font=\footnotesize}}

\nextgroupplot[
	ylabel={\footnotesize FID},
    ytick={0, 5, 10, 20, ..., 50},
]

\addplot [thick, color=myorange] coordinates {
	(0, 2.6721683805157)
	(55, 2.6721683805157)
};

\addplot [thick, mark=halfsquare right*, color=mygreen2] table[x=QF, y=FID] {fig/imagenet_methods/qgac_regression.csv};

\addplot [thick, mark=halfsquare left*, color=mygreen1] table[x=QF, y=FID] {fig/imagenet_methods/qgac_merp.csv};

\addplot [thick, mark=square*, color=myred2] table[x=QF, y=FID] {fig/imagenet_methods/bahat.csv};

\addplot [thick, mark=pentagon*, color=myyellow] table[x=QF, y=FID] {fig/imagenet_methods/fbcnn.csv};

\addplot [thick, mark=diamond*, color=myred1] table[x=QF, y=FID] {fig/imagenet_methods/swinir.csv};

\addplot [thick, mark=*, color=myblue2] table[x=QF, y=FID] {fig/imagenet_methods/ours.csv};

\addplot [thick, mark=triangle*, mark options={solid}, color=myblue1, dashed] table[x=QF, y=FID] {fig/imagenet_methods/ours_p.csv};

\nextgroupplot[
	ymax=4.2,
	ylabel={\footnotesize Consistency-RMSE},
    ytick={0, ..., 10},
]

\addplot [thick, color=myorange] coordinates {
	(0, 0)
	(55, 0)
};
\label{GT}

\addplot [thick, mark=halfsquare right*, color=mygreen2] table[x=QF, y=RMSE] {fig/imagenet_methods/qgac_regression.csv};
\label{QGAC}

\addplot [thick, mark=halfsquare left*, color=mygreen1] table[x=QF, y=RMSE] {fig/imagenet_methods/qgac_merp.csv};
\label{QGAC-GAN}

\addplot [thick, mark=pentagon*, color=myyellow] table[x=QF, y=RMSE] {fig/imagenet_methods/fbcnn.csv};
\label{FBCNN}

\addplot [thick, mark=diamond*, color=myred1] table[x=QF, y=RMSE] {fig/imagenet_methods/swinir.csv};
\label{SwinIR}

\addplot [thick, mark=square*, color=myred2] table[x=QF, y=RMSE] {fig/imagenet_methods/bahat.csv};
\label{Bahat}

\addplot [thick, mark=*, color=myblue2] table[x=QF, y=RMSE] {fig/imagenet_methods/ours.csv};
\label{Ours}

\addplot [thick, mark=triangle*, mark options={solid}, color=myblue1, dashed] table[x=QF, y=RMSE] {fig/imagenet_methods/ours_p.csv};
\label{Ours-P}

\nextgroupplot[
	ylabel={\footnotesize PSNR},
    ytick={1, 3, ..., 40},
]

\addplot [thick, mark=diamond*, color=myred1] table[x=QF, y=PSNR] {fig/imagenet_methods/swinir.csv};

\addplot [thick, mark=diamond*, color=myyellow] table[x=QF, y=PSNR] {fig/imagenet_methods/fbcnn.csv};

\addplot [thick, mark=halfsquare right*, color=mygreen2] table[x=QF, y=PSNR] {fig/imagenet_methods/qgac_regression.csv};

\addplot [thick, mark=halfsquare left*, color=mygreen1] table[x=QF, y=PSNR] {fig/imagenet_methods/qgac_merp.csv};

\addplot [thick, mark=square*, color=myred2] table[x=QF, y=PSNR] {fig/imagenet_methods/bahat.csv};

\addplot [thick, mark=*, color=myblue2] table[x=QF, y=PSNR] {fig/imagenet_methods/ours.csv};

\addplot [thick, mark=triangle*, mark options={solid}, color=myblue1, dashed] table[x=QF, y=PSNR] {fig/imagenet_methods/ours_p.csv};

\end{groupplot}

\matrix [
 matrix of nodes,
 nodes={anchor=north},
 anchor=north,
 at={([shift={(0,-1)}]$(group c1r1.south)!0.5!(group c3r1.south)$)},
 fill=white,
 draw,
 inner sep=2pt,
 row sep=2pt,
 column sep=4pt,
] {
\scriptsize{\ref{GT} Ground-Truth} &
\scriptsize{\ref{QGAC} QGAC~\cite{QuantizationGuidedJPEG2020ehrlich}} &
\scriptsize{\ref{QGAC-GAN} QGAC-GAN~\cite{QuantizationGuidedJPEG2020ehrlich}} &
\scriptsize{\ref{SwinIR} SwinIR~\cite{SwinIRImageRestoration2021lianga}} &
\scriptsize{\ref{FBCNN} FBCNN~\cite{FlexibleBlindJPEG2021jiang}} &
\scriptsize{\ref{Bahat} Bahat~\cite{WhatImageExplorable2021bahat}} &
\scriptsize{\ref{Ours} Ours} &
\scriptsize{\ref{Ours-P} Ours-P} \\
};

\end{scope}
\node (title) at ($(group c1r1.north)!0.5!(group c3r1.north)$) [above, yshift=\pgfkeysvalueof{/pgfplots/every axis title shift}] {\large ImageNet-ctest10k};

\end{tikzpicture}
\caption{FID, Consistency and PSNR results of different methods on ImageNet-ctest10k across multiple QFs. By using explicit control over the perception-consistency tradeoff and by avoiding the perception-distortion tradeoff, our method provides SoTA FID results across a wide-range of QFs while being consistent with the compressed inputs.
Each method is evaluated only on QFs it was trained on.
The consistency of Ours-P and Bahat are practically zero -- see \Cref{sec:experiments-results-ffhq} for more details.}
\label{fig:imagenet_methods}
\end{figure*}
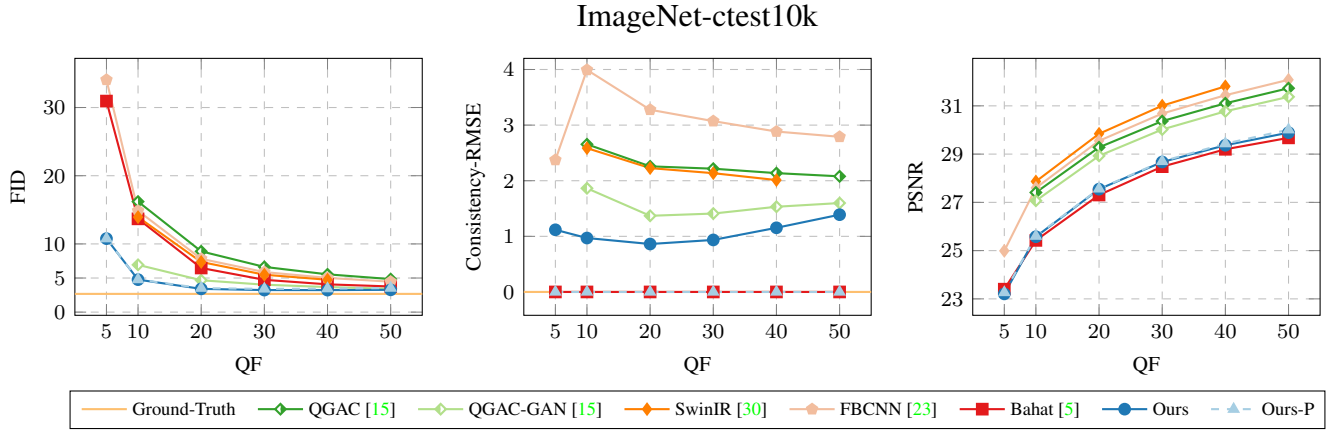

\noindent {\bf Perceptual quality \& consistency:}
Following the trends shown on FFHQ-128, Ours-P creates consistent reconstructions across a wide range of QFs similarly to Bahat's, but with much improved perceptual quality.

When comparing to QGAC, SwinIR and FBCNN we see a predictable result -- the regression models achieves the worst perceptual quality and QGAC-GAN model shows better results compared to Bahat's model, but they all fail to generate consistent reconstructions. This is a direct result of the perception-distortion tradeoff~\cite{PerceptionDistortionTradeoff2018blau} as also manifested in the PSNR results.

While QGAC, SwinIR and FBCNN are not trained to minimize an MSE loss, they achieve SoTA results in terms of PSNR on JPEG reconstruction, hence, they could be seen as a compelling MMSE estimator candidates. As such, they should have near-perfect consistency following \autoref{thm:mmse}, but clearly this is not the case. Similarly to the experiments shown in \Cref{sec:experiments-results-ffhq} on Ours-A and Ours\=/MSE, it seems that explicit supervision on the consistency of the reconstructed images, or enforcing it via the architecture, is necessary to achieve consistent results.

\subsection{Ablation}
\label{sec:ablation}

As explained in \Cref{sec:practice.train} our method is trained using several loss terms. For the FFHQ-128 data set we use a GAN loss, a consistency loss, and the first-moment penalty, while for ImageNet we also use a VGG ``perceptual'' loss and a second-moment penalty. In \autoref{tab:ablation} we present an ablation study of training the model with the different terms on the ImageNet data set, and here we provide a detailed explanation for each row in the table:
{\bf Baseline:} By using GAN and consistency losses alone (\autoref{eq:uncons_opt}) we attain better perceptual quality (FID) than Bahat~\cite{WhatImageExplorable2021bahat} but with lower variability (Per-Pixel STD);
{\bf $+$ First-Moment:} As noted in \cite{HighPerceptualQuality2021ohayon}, the first-moment  penalty alleviates the mode-collapse issue of conditional GANs, and indeed, it significantly improves the output variation of our model;
{\bf $+$ VGG:} A perceptual loss further improves the perceptual quality at the cost of lower variability;
{\bf $+$ Second-Moment:} The new penalty increases output variation without hindering perceptual quality, hence suggesting that the increased output variation is of meaningful details;
{\bf $+$ Ours:} With increased second-moment coefficient we further improve the output variability;
{\bf $+$ Ours-P:} By enforcing perfect consistency via projection we still attain much better perceptual quality and output variation compared to Bahat~\cite{WhatImageExplorable2021bahat};
{\bf Perceptual Baseline:} Without explicitly requiring consistency we achieve similar FID to our method but without consistency with the measurements.

\begingroup
\setlength{\tabcolsep}{3pt} %
\begin{table*}[]
    \rowcolors{2}{gray!10}{white}
    \centering
    \caption{Ablation study on the ImageNet-ctest10k dataset at QF=10. The VGG loss is crucial for good-perceptual quality without hurting consistency. The second-moment penalty increases per-pixel standard deviation (a value between 0 and 1) without hurting either perceptual quality nor consistency. The consistency of constrained methods, marked by *, are practically zero -- see \Cref{sec:experiments-results-ffhq} for more details.}
    \begin{tabular}{|l|c|c|c|c|c||c|c|c|c|}
	\hline
	\rowcolor{gray!25}
	Method &
	GAN &
	C &
	FM &
	P &
	SM &
	
	FID {($\downarrow$)} &
	Consistency {($\downarrow$)} &
	PSNR {($\uparrow$)} &
	Per-Pixel STD {($\uparrow$)} \\
	
	Perceptual Baseline & \checkmark & & \checkmark & \checkmark & & $4.54$ & 6.9724 & 26.0612 & 0.0111 \\
	Baseline & \checkmark & \checkmark & & & &  $9.12$ & 0.9650 & 25.5061 & 0.0021 \\
	$+$ First-Moment & \checkmark & \checkmark & \checkmark & & &  $8.83$ & 0.9694 & 25.5767 & 0.0064 \\
	$+$ VGG & \checkmark & \checkmark & \checkmark & \checkmark & & $4.71$ & 0.9411 & 25.8906 & 0.0026 \\
	$+$ Second-Moment & \checkmark & \checkmark & \checkmark & \checkmark & \checkmark $(10^1)$ & $4.70$ & 0.9452 & 25.8539 & 0.0063 \\
	Ours & \checkmark & \checkmark & \checkmark & \checkmark & \checkmark $(10^3)$ & $4.76$ & 0.9696 & 25.5758 & 0.0194 \\
	Ours-P & \checkmark & \checkmark & \checkmark & \checkmark & \checkmark $(10^3)$ & $4.78$ & $\approx0$* & 25.6008 & 0.0188 \\

	\hline\hline
	\rowcolor{gray!25}
	Bahat & \multicolumn{5}{c||}{} & $13.71$ & $\approx0$* & 25.4243 & 0.0040 \\
	
	\hline
	\end{tabular}
	\label{tab:ablation}
\end{table*}
\endgroup

\begingroup
\setlength{\tabcolsep}{0pt} %
\renewcommand{\arraystretch}{0} %

\setlength{\hero}{0.3333\columnwidth}

\newcommand{\addimgcol}[7][1]{
	\centered{
	   	\begin{tikzpicture}[
	   		baseline=-2.45,
	   		spy using outlines={magnification=#3, circle, height=1.5cm, width=1.5cm, yellow, every spy on node/.append style={thick}, connect spies},
	   		]
			\node[inner sep=0pt]{\scalebox{#1}[1]{\adjincludegraphics[width=\hero, trim={#4}, clip]{images/ours/#7-imagenet/compressed/#2}}};
			\spy on (#5) in node at (#6);
		\end{tikzpicture}}&
		
		\centered{
		\begin{tikzpicture}[
	   		baseline=-2.45,
	   		spy using outlines={magnification=#3, circle, height=1.5cm, width=1.5cm, yellow, every spy on node/.append style={thick}, connect spies},
	   		]
			\node[inner sep=0pt]{\scalebox{#1}[1]{\adjincludegraphics[width=\hero, trim={#4}, clip]{images/qgac/#7-imagenet/#2}}};
			\spy on (#5) in node at (#6);
		\end{tikzpicture}}&
		
		\centered{
		\begin{tikzpicture}[
	   		baseline=-2.45,
	   		spy using outlines={magnification=#3, circle, height=1.5cm, width=1.5cm, yellow, every spy on node/.append style={thick}, connect spies},
	   		]
			\node[inner sep=0pt]{\scalebox{#1}[1]{\adjincludegraphics[width=\hero, trim={#4}, clip]{images/qgac-gan/#7-imagenet/#2}}};
			\spy on (#5) in node at (#6);
		\end{tikzpicture}}&
		
		\centered{
	   	\begin{tikzpicture}[
	   		baseline=-2.45,
	   		spy using outlines={magnification=#3, circle, height=1.5cm, width=1.5cm, yellow, every spy on node/.append style={thick}, connect spies},
	   		]
			\node[inner sep=0pt]{\scalebox{#1}[1]{\adjincludegraphics[width=\hero, trim={#4}, clip]{images/bahat/#7-imagenet/fake_0/#2}}};
			\spy on (#5) in node at (#6);
		\end{tikzpicture}}&

		\centered{
		\begin{tikzpicture}[
	   		baseline=-2.45,
	   		spy using outlines={magnification=#3, circle, height=1.5cm, width=1.5cm, yellow, every spy on node/.append style={thick}, connect spies},
	   		]
			\node[inner sep=0pt]{\scalebox{#1}[1]{\adjincludegraphics[width=\hero, trim={#4}, clip]{images/ours-p/#7-imagenet/fake_0/#2}}};
			\spy on (#5) in node at (#6);
		\end{tikzpicture}}
		
	    \\
}

\begin{figure*}[t]
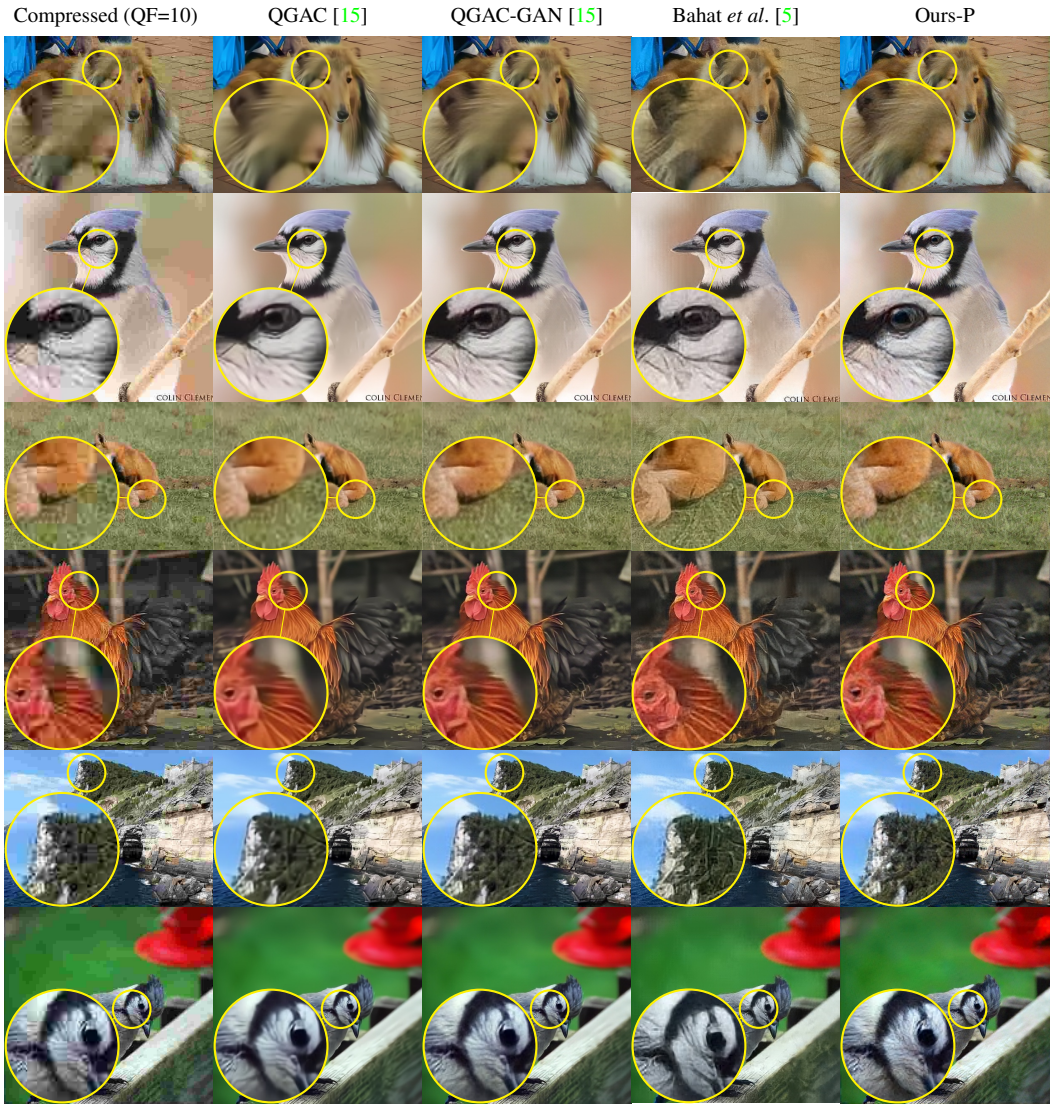

	\centering
    \begin{tabular}{c c c c c}

	    \footnotesize{Compressed (QF=10)} &
	    \footnotesize{QGAC~\cite{QuantizationGuidedJPEG2020ehrlich}} &
	    \footnotesize{QGAC-GAN~\cite{QuantizationGuidedJPEG2020ehrlich}} &
	    \footnotesize{Bahat~\etal~\cite{WhatImageExplorable2021bahat}} &
	    \footnotesize{Ours-P}
	    \\
	    
	   	\rule{0pt}{0.8ex}\\
	   	
	   	\addimgcol{ILSVRC2012_val_00038789}{3}{0 0 0 0}{-0.1,0.6}{-0.625,-0.275}{10}
	   	
	   	\addimgcol{ILSVRC2012_val_00041308}{3}{{.05\width} {.05\height} {.05\width} {.05\height}}{-0.15,0.65}{-0.625,-0.625}{10}
	    
	    \addimgcol{ILSVRC2012_val_00042670}{3}{0 0 0 0}{0.5,-0.3}{-0.625,-0.225}{10}
	    
	    \addimgcol{ILSVRC2012_val_00038826}{3}{0 {.1\height} 0 {.1\height}}{-0.4,0.8}{-0.625,-0.56}{10}
	    
	    \addimgcol{ILSVRC2012_val_00039874}{3}{0 0 0 0}{-0.3,0.75}{-0.625,-0.27}{10}
	    
	    \addimgcol[-1]{ILSVRC2012_val_00034552}{3}{0 0 0 0}{0.3,-0.05}{-0.625,-0.54}{10}
	    
    \end{tabular}
    \caption{Reconstruction examples on ImageNet-ctest10k at QF=10; Bahat is the only other method that generates consistent reconstructions and aims for high perceptual quality;
    QGAC-GAN's reconstructions are sharp but not consistent, hence could not have created the input image;
    Our method generates finer details while still being consistent, as seen in \autoref{fig:imagenet_methods};
    Zoom-in on interesting regions are shown.}
	\label{fig:imagenet_compare}
\end{figure*}
\endgroup

\section{Summary}

In this work we approach the JPEG decompression task from an uncommon direction -- generate visually pleasing and consistent reconstructions by leveraging recent advancements in image restoration, such as the perception-distortion and the perception-consistency theoretical tradeoffs. Using these tools we surpass prior work and provide decompressed JPEG images of tunable consistency and high perceptual quality.

The proposed solution is based on a stochastic conditional GAN with carefully tailored loss function that promotes detailed and vivid results, consistency to the measurements, proximity to the training data without scarifying quality, and a spread of the randomized results. Our future work will focus on better diversifying the obtained solutions and on a quest for a tractable computational method for evaluating the proximity of the obtained model to the ideal posterior sampler.

\textbf{Acknowledgement}
This research was partially supported by the Israel Science Foundation (ISF) under Grant 335/18 and the Council For Higher Education - Planning \& Budgeting Committee.

\appendix

\clearpage

\section{Implementation details}
\label{sec:imp}

\subsection{Architecture}
\label{sec:imp-arch}

Our GAN architecture is composed of an RRDB based generator and a ResNet critic.
The generator largely follows the architecture proposed in Real-ESRGAN~\cite{RealESRGANTrainingRealWorld2021wanga, basicsr}, where we use a pixel-unshuffle block to rearrange a $m \times n \times 3$ input into a $m/2 \times n/2 \times 12$ input to reduce the computational complexity of the following 23 RRDB blocks, followed by an upscale of the result to restore the original input shape. One major difference from \cite{RealESRGANTrainingRealWorld2021wanga} is the injection of noise channels along the network -- Each ``Noise Injection'' block concatenates a new channel filled with Gaussian noise, which is used by the network to generate stochastic details (this is $Z$, mentioned in the loss formulation). 
In cases supporting multiple QFs we implement a FiLM~\cite{FiLMVisualReasoning2018perez} block that modulates the results of an RRDB or upsampling block using a learned affine transformation, conditioned on the quantization table that is embedded in the JPEG file. We present the generator architecture in \autoref{fig:generator}.
The critic is a plain ResNet34~\cite{DeepResidualLearning2016he}. 

\subsection{Training}
\label{sec:imp-train}

We train our models using the Adam~\cite{AdamMethodStochastic2015kingmaa} optimizer with batch-size of 32 to alternately update the generator and the critic networks, using \autoref{eq:opt_g} and \autoref{eq:opt_d}. We use exponential moving average on the generator weights with decay factor of $0.999$.

\noindent {\bf FFHQ-128:}
The learning rate starts at $1 \times 10^{-4}$ and annealed to $1 \times 10^{-7}$ after $400,000$ steps using cosine annealing.
We scale $V(D_\omega, G_\theta)$ in \autoref{eq:opt_g} by $1 \times 10^{-3}$ and set $\lambda_{R_1}=1$, $\lambda_{FM}=1$ and $\lambda_{C}$ according to the version reported in \autoref{fig:ffhq_methods}. For the versions denoted as {\bf Ours} and {\bf Ours\=/P} we use cosine annealing of $\lambda_{C}$ from $1 \times 10^{-1}$ to $1 \times 10^{1}$.

\noindent {\bf General-Content:}
The learning rate starts at $1 \times 10^{-4}$ and annealed to $1 \times 10^{-6}$ after $400,000$ steps using cosine annealing. The weights are initialized from a model trained for $50,000$ steps as a regression model using an MSE loss, as this was found to be important for stabilizing the training.
We scale $V(D_\omega, G_\theta)$ in \autoref{eq:opt_g} by $1 \times 10^{-3}$ and set $\lambda_{R_1}=1$, $\lambda_{FM}=1$, $\lambda_{P}=1 \times 10^{-2}$, $\lambda_{SM}=1 \times 10^{3}$ and use cosine annealing of $\lambda_{C}$ from $1 \times 10^{0}$ to $1 \times 10^{3}$.

To estimate the mean and variance of generated images for use in $FM(G_\theta)$ and $SM(G_\theta)$ we generate 16 different restorations (using different $Z$s) for each of the first 8 images in a batch. In order to save train time we preform this every 8 iterations as we found negligible performance differences compared to performing this each iteration.

As the reference MMSE estimator for $SM(G_\theta)$ we use a regression model with the same architecture, trained for $650,000$ steps using a simple MSE loss.

As training data, we extract square patches with random scale (between $128 \times 128$ and the image resolution) from the training set at random position and rescale them to $128 \times 128$ pixels. This is inspired by \cite{AnyResolutionTrainingHighResolution2022chai} and it exposes our GAN to more diverse set of patches.

\begingroup

\begin{figure*}
	\centering
	
	\includegraphics[width=\textwidth]{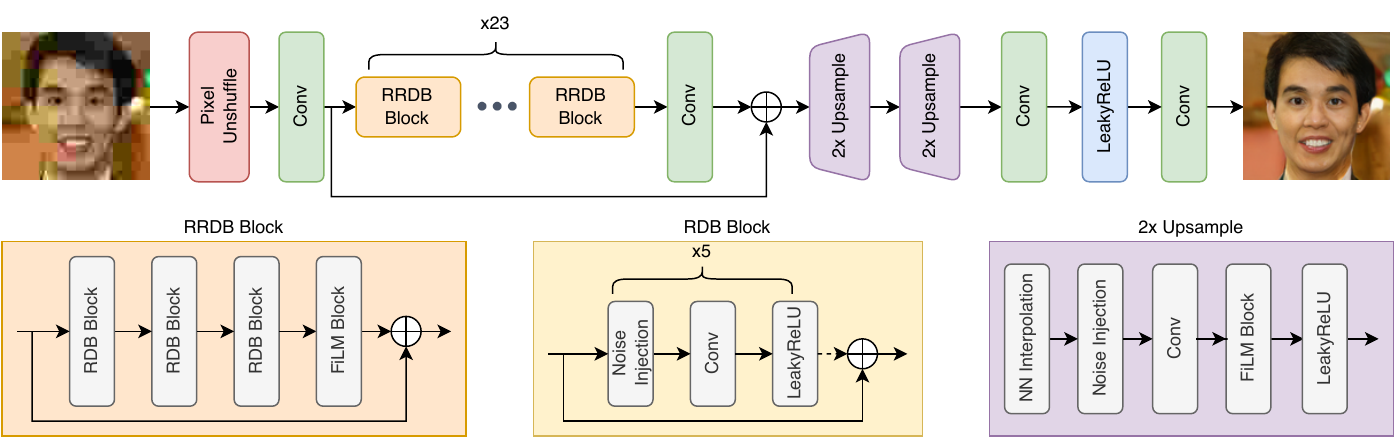}    

	\caption{Our generator's architecture is based on \cite{RealESRGANTrainingRealWorld2021wanga}, where Residual-in-Residual Dense Blocks (RRDBs) are used to restore a degraded input. Due to the fact that we work on full resolution inputs, we trade spatial resolution with depth using a pixel-unshuffle block that rearrange the input from $m \times n \times 3$ tensor into a $m/2 \times n/2 \times 12$ tensor and in the end we upsample the resulted tensor. To achieve stochasticity, we inject a noise channel to the features at the beginning of every RDB block and in the upsampling blocks. To be agnostic to the QF of the input image, we condition the RRDB and upsampling blocks on the quantization table used to create the image using a FiLM~\cite{FiLMVisualReasoning2018perez} block.}
	\label{fig:generator}
\end{figure*}

\endgroup

\section{MMSE estimator is consistent}
\label{sec:mmse_proof}

\printProofs

\section{Second-moment penalty}
\label{sec:second_moment}

The \textbf{per-pixel} conditional variance $\sigma_{X|Y}^2$ of a random-variable $X$ can be estimated from samples $\{x_i\}_{i=1}^n$ sampled from $p_{X|Y}$:
\begin{equation}
    \begin{aligned}
        \sigma_{X|Y}^2 \approx \frac{1}{n} \sum_{i=1}^n (x_i - \mu_{X|Y})^2,
    \end{aligned}
\end{equation}
where $\mu_{X|Y}$ is the mean of $X$ conditioned on $Y$. In practice we do not have access to this mean, hence we can approximate it either using an MMSE estimator $\bar{X}(Y)$ (that at optimality becomes the conditional mean) or using the sample mean:
\begin{equation}
    \begin{aligned}
        \tilde\mu_{X|Y} = \frac{1}{n} \sum_{i=1}^n x_i.
    \end{aligned}
\end{equation}

In our case, we have two different variables for which we would like to compare their variances -- the ground-truth images $X$ and our reconstructed images $\hat{X}$, both conditioned on the compressed images $Y$.

Recall that for a fixed $Y$ we have only a single ground-truth sample $X$ and thus we cannot compute a useful sample conditional mean, hence we opt to use a pre-trained regression model as our MMSE estimator $\bar{X}(Y)$:
\begin{equation}
    \begin{aligned}
        \tilde\sigma_{X|Y}^2 = (x - \bar{X}(Y))^2.
    \end{aligned}
\end{equation}
Intuitively, $\bar{X}(Y)$ averages all possible values of each pixel in the reconstructed image based on the probability of the value. Hence, pixels with similar values in $\bar{X}(Y)$ and in $x$ are pixels with small ambiguity -- all of the possible reconstructions agree on the pixel's value, and thus we can expect small variance at such locations. 
On the other hand, pixels with large discrepancy between $\bar{X}(Y)$ and $x$ are not necessarily an indication for large variance as the value in $x$ might be rare and thus far from the mean. To better analyze the latter case we need further assumption on the conditional probability $X|Y$.

As $p_{\hat{X}|Y}$ changes during training, we cannot use a pre-trained regression model that was trained on $p_{X|Y}$, but we can generate as much samples as needed, hence we can compute a sample conditional mean that approximates well the true mean:
\begin{equation}
    \begin{aligned}
        \tilde\sigma_{\hat{X}|Y}^2 = \frac{1}{n} \sum_{i=1}^n (\hat{x}_i - \tilde\mu_{\hat{X}|Y})^2,
    \end{aligned}
\end{equation}
To create the penalty mentioned in \Cref{sec:practice.train} we just compute the distance between those two, per-pixel, variance approximations over a batch of realizations of $Y$:
\begin{equation}
    \begin{aligned}
        SM&(\hat{X}) = \lambda_{SM} \mathbb{E}_{X,\hat{X},Y} \left[ \norm{ \tilde\sigma_{X|Y}^2 - \tilde\sigma_{\hat{X}|Y}^2} \right].
    \end{aligned}
\end{equation}

In \autoref{fig:imagenet_var} we present a visual illustration of the variance estimator $\tilde\sigma_{X|Y}^2$ using QGAC as our MMSE estimator alongside per-pixel sample variance of Ours' and Bahat~\etal's methods.

\begingroup
\newcolumntype{M}[1]{>{\centering\arraybackslash}m{#1}}
\setlength{\tabcolsep}{0pt} %
\renewcommand{\arraystretch}{0} %

\setlength{\hero}{0.2\textwidth}

\begin{figure*}[t]
	\centering
	\begin{tabular}{c}
	\begin{tabular}{c M{0.05\textwidth}}
    \begin{tabular}{c c c c c}
    	Compressed ($Y$) &
    	QGAC ($\bar{X}(Y)$) &
    	Ground-Truth ($X$) &
    	$(X-\bar{X}(Y))^2$ \\
    	\rule{0pt}{0.8ex}\\
    	
    	\noisy[\hero]{10}{imagenet}{ILSVRC2012_val_00041478}&
    	\qgac[\hero]{10}{imagenet}{ILSVRC2012_val_00041478}&
    	\real[\hero]{10}{imagenet}{ILSVRC2012_val_00041478}&
    	\centered{\includegraphics[width=\hero]{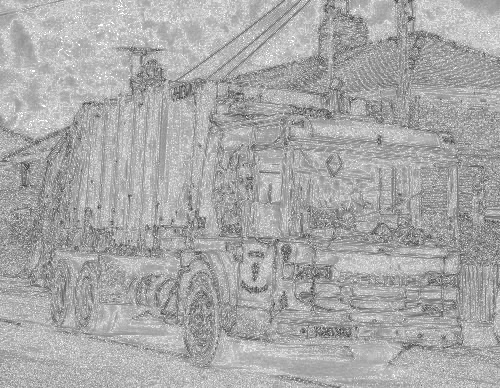}}
    	\\
    	
    \end{tabular}
    &
    \vspace{.25cm}
        \begin{tikzpicture}
            \begin{axis}[
                hide axis,
                scale only axis,
                height=2.5cm,
                width=1cm,
                colormap = {whiteblack}{color(0cm)  = (white);color(1cm) = (black)},
                colorbar,
                point meta min=0,
                point meta max=1,
                colorbar style={
                }]
                \addplot [draw=none] coordinates {(0,0)};
            \end{axis}
            \end{tikzpicture}
    \end{tabular}
    
    \\
   	\rule{0pt}{0.8ex}
   	\\
    
    \begin{tabular}{c c c c c}
	    Bahat~\etal & Baseline & + VGG & + SM & Ours \\
	   	\rule{0pt}{0.8ex}\\
	   	
	    \bahats[\hero]{10}{imagenet}{ILSVRC2012_val_00041478}&
	    \centered{\includegraphics[width=\hero]{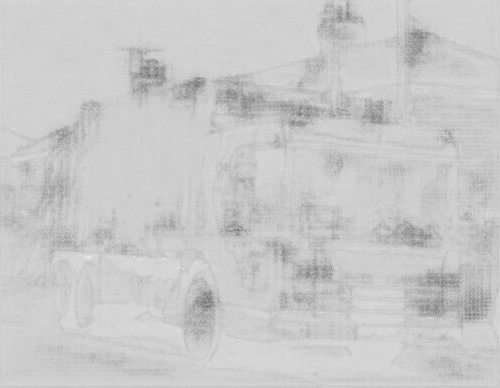}}&
	    \centered{\includegraphics[width=\hero]{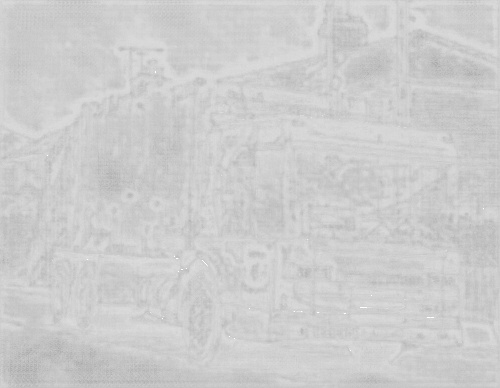}}&
	    \centered{\includegraphics[width=\hero]{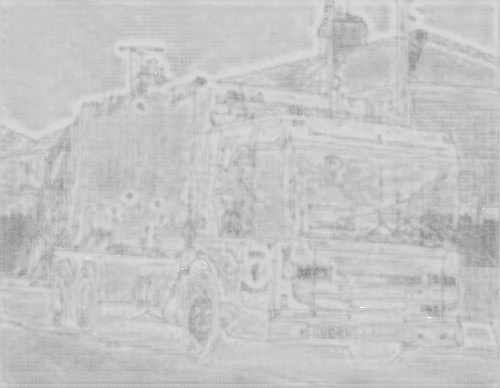}}&
	    \ourss[\hero]{10}{imagenet}{ILSVRC2012_val_00041478}	    
    \end{tabular}
    \end{tabular}
	\caption{
	We approximate the conditional variance of clean images on a compressed input ($Y$) using a single ground-truth image ($X$) and an MMSE estimator ($\bar{X}(Y)$).
	Here we use QGAC~\cite{QuantizationGuidedJPEG2020ehrlich} as our MMSE estimator on a JPEG compressed image with QF=10. On the bottom row we present per-pixel variance maps calculated on 64 samples of different methods ($\hat{X}$). 
	For visualization purposes we use the 8th root of the variance and add a color bar (white and black correspond to low and high variance, respectively).
	The variance estimation $(X - \bar{X}(Y))^2$ indicates that we should expect more variance in regions with sharp transitions and this is indeed what we see in the variance maps of the different methods.
	Our method produce more variance compared to Bahat~\etal's method in regions with sharp transition and less variance in smooth regions.
	This is also supported quantitatively in \autoref{tab:ablation} where we see larger average per-pixel standard deviation while achieving better FID.
	}
	\label{fig:imagenet_var}
\end{figure*}
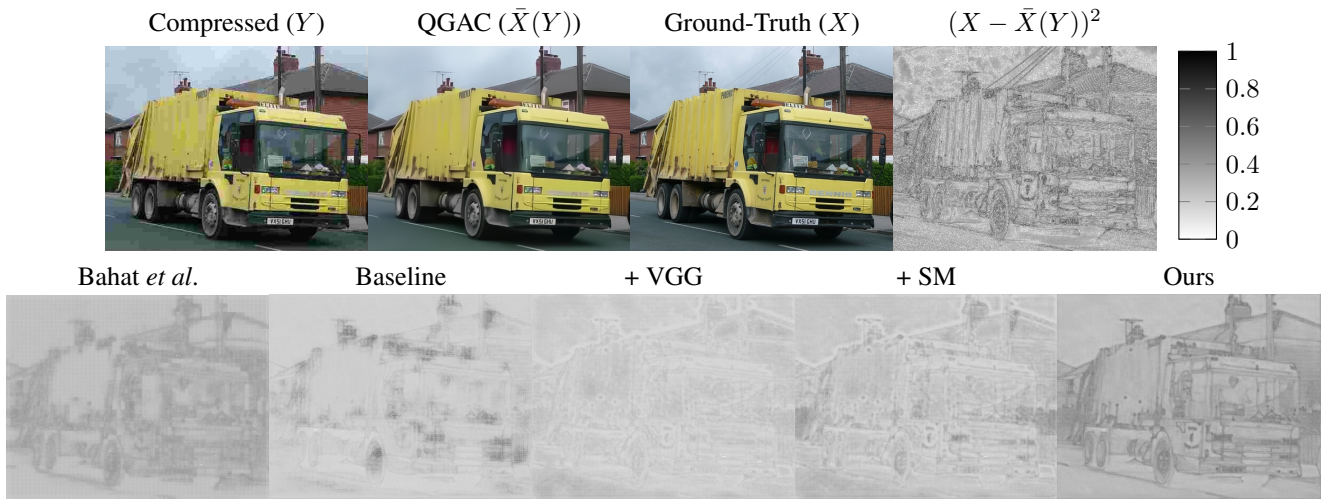
\endgroup

\section{JPEG numerical errors}
\label{sec:numerical-erros}

As mentioned in \Cref{sec:experiments-results-ffhq} and \autoref{tab:ffhq}, numerical approximations in the JPEG algorithm result in near, but not perfect, consistent results. To showcase that this phenomena is not unique to the differentiable JPEG implementations used by us and \cite{WhatImageExplorable2021bahat} we test \emph{libjpeg-9d}~\cite{IndependentJPEGGroup}, a standard JPEG implementation used in packages such as OpenCV~\cite{opencv_library} and PIL~\cite{clark2015pillow}.
We compress the FFHQ test set with QF=100 which corresponds to no quantization, at block size of 1 which means the DCT values equal the color values, and without chroma sub-sampling, meaning we should expect a perfect reconstruction of the inputed images theoretically. \autoref{tab:libjpeg} present the RMSE between the ground-truth images and the compressed images, using different configurations of the encoder-decoder.
As it can be clearly seen, as long as we do not skip the YCbCr color-space conversion, true lossless compression is not achieved due to numerical approximations. \autoref{tab:consistency} presents the actual Consistency RMSE results of the projected methods.
We can see that while all the methods are not perfectly consistent, they are extremely quite to zero and perform better than \emph{libjpeg-9d} when operating at the YCbCr color space.

To reproduce the \emph{libjpeg-9d} images, use the following snippet:

\begin{lstlisting}[language=bash]
# 2D-DCT=float, Color-Space=YCbCr, Block-Size=1
cjpeg -block 1 -quality 100 -sample 1x1,1x1,1x1 -dct float <input_image> | djpeg -dct float > <output_image>

# 2D-DCT=float, Color-Space=RGB, Block-Size=1
cjpeg -block 1 -quality 100 -sample 1x1,1x1,1x1 -dct float -rgb <input_image> | djpeg -dct float > <output_image>
\end{lstlisting}

\begin{table}
    \rowcolors{2}{gray!10}{white}
    \centering
    \caption{RMSE between clean FFHQ test images and compressed-decompressed FFHQ test images using the \emph{libjpeg-9d} library. We use no chroma sub-sampling and quantization table of ones, hence, the process should be invertible. In practice, we see deviations due to numerical approximations.}
    \begin{tabular}{|c|c|c||c|}
	\hline
	\rowcolor{gray!25}
	2D-DCT & Color-Space & Block-Size & RMSE \\
	float & YCbCr & 1 & 0.5465 \\
	float & RGB & 1 & 0 \\
	\hline
	\end{tabular}
	\label{tab:libjpeg}
\end{table}

\begin{table}
    \rowcolors{2}{gray!10}{white}
    \centering
    \caption{Consistency results of different constrained methods on the FFHQ test dataset at QF=5. The inconsistencies stem from numerical approximations.}
    \begin{tabular}{|l||c|}
	\hline
	\rowcolor{gray!25}
	Method & Consistency \\
	Bahat			& 0.0867 \\
	$\lambda_C{=}0$-P		& 0.9466 \\
	Ours-P		& 0.1664 \\
	\rowcolor{gray!25}
	Ground Truth (theoretical)         & 0 \\
	\hline
	\end{tabular}
	\label{tab:consistency}
\end{table}

\section{More results}
\label{sec:more_results}

\subsection{FFHQ}

In \autoref{tab:ffhq} we present quantitative results of the different methods on the FFHQ test set at QF=5.

In \autoref{fig:ffhq_zoom} we visually demonstrate the perceptual quality and the stochastic variation of our method by presenting several realizations of a given input and comparing them to the other methods. 
As expected, the regression model generates overly-smoothed results and is unable to recover fine details such as hairs and wrinkles.
Bahat's method successfully recovers some fine details but suffers from severe color and grid-like artifacts. We find that the training of this method is highly unstable, and we hypothesize that this is partly due to the overly constrained optimization with perfect consistency.
The results denoted Ours are highly visually appealing -- fine details such as hair and wrinkles are generated in a reasonable manner.

In \autoref{fig:ffhq_recompress} we present a couple of compressed images from the test set of FFHQ and the corresponding recompressed restoration from our method with and without consistency regularization and with projection. This showcases the effectiveness of our consistency regularization in improving the consistency of the reconstructed images without detoriating their perceptual quality.

In \autoref{fig:ffhq_variation} we present the stochastic nature of the different methods (except the deterministic regression models). We expect to see different plausible details generated for the same compressed input image $Y$ given different noise injection $Z$. Indeed, we see that our methods generate slight variations in the expression, in the beard and hair structure, in the background details and in the skin colors. Those variations are also indicated by the per-pixel standard deviation map we present for each method, where darker values represent more varying pixels in the restored images. While Bahat's method also produces stochastic results, it can be clearly seen that most of the variation comes from color artifacts.

In \autoref{fig:ffhq_perceptual} we present more results of the different methods on the test set of FFHQ.

\begingroup
\setlength{\tabcolsep}{3pt} %
\begin{table}[h]
    \rowcolors{2}{gray!10}{white}
    \centering
    \caption{Quantitative results of different methods on the FFHQ test set at QF=5. The consistency of constrained methods, marked by *, are practically zero -- please refer to \Cref{sec:experiments-results-ffhq} and \Cref{sec:numerical-erros} for more details.}
    \begin{tabular}{|l||c|c|c|}
	\hline
	\rowcolor{gray!25}
	Method &         FID {($\downarrow$)} & Consistency {($\downarrow$)} & PSNR {($\uparrow$)} \\
	Regression        		&  $66.14 \pm 0.00$ & 5.2217 & 25.6579 \\
	Ours-A        		&  $35.26 \pm 0.00$ & 0.3203 & 25.4529 \\
	
	Bahat			& $65.86 \pm 0.28$ & $\approx0$* & 22.6807 \\
	
	$\lambda_C{=}0$        		&  $16.46 \pm 0.16$ & 11.8848 & 23.3457 \\
	$\lambda_C{=}0$-P		& $20.60 \pm 0.16$ & $\approx0$* & 23.4896 \\
	Ours        		&  $16.70 \pm 0.14$ & 0.7481 & 23.7595 \\
	Ours-P		& $18.54 \pm 0.14$ & $\approx0$* & 23.7767 \\
	
	\rowcolor{gray!25}
	Ground Truth         &  $10.28 \pm 0.00$ & 0 & $\infty$ \\
	\hline
	\end{tabular}
	\label{tab:ffhq}
\end{table}
\endgroup

\subsection{ImageNet}

In \autoref{tab:imagenet} we present quantitative results of the different methods on ImageNet-ctest10k at QF=10.

In \autoref{fig:imagenet_5_perceptual} and \autoref{fig:imagenet_10_perceptual} we present more results of the different methods on ImageNet-ctest10k. Note that QGAC and QGAC-GAN are not trained on QFs lower than 10, hence we do not show their results on QF=5 for fair comparison. The visual results further corroborate the quantitative results shown in \autoref{fig:imagenet_methods} -- Our method provides the best perceptual results, creating more fine details and less artifacts and projecting our results does not detoriate their perceptual quality. Note that while QGAC-GAN provide better visual results compared to Bahat's, they are not consistent with the compressed inputs. This means that those are not valid reconstructions in the sense that they could not have created the compressed images.

\subsection{LIVE1 \& BSDS500}

In \autoref{fig:live1_perceptual} and \autoref{fig:bsds_perceptual} we present visual results of different methods on LIVE1~\cite{LIVEImageQualitysheikh, StatisticalEvaluationRecent2006sheikh} and BSDS500 \cite{ContourDetectionHierarchical2011arbelaez} datasets.

Such small datasets (29 and 500 images, respectively) cannot be used for reliable deep-features-based, ensemble perceptual quality assessments (FID, KID, IS, etc.). From the official FID implementation\footnote{\url{https://github.com/bioinf-jku/TTUR}} ``IMPORTANT: The number of samples [...] should be greater than the dimension of the coding layer, here 2048 [...]''.
Hence, we do not include quantitative results for this datasets. 

\begingroup
\setlength{\tabcolsep}{3pt} %
\begin{table}[h]
    \rowcolors{2}{gray!10}{white}
    \centering
    \caption{Quantitative results of different methods on ImageNet-ctest10k at QF=10. The consistency of constrained methods, marked by *, are practically zero -- please refer to \autoref{sec:experiments-results-ffhq} and \autoref{sec:numerical-erros} for more details.}
    \begin{tabular}{|l||c|c|c|}
	\hline
	\rowcolor{gray!25}
	Method &         FID {($\downarrow$)} & Consistency {($\downarrow$)} & PSNR {($\uparrow$)} \\
	SwinIR			& $13.93 \pm 0.00$ & 2.5858 & 27.8662 \\
	FBCNN			& $14.85 \pm 0.00$ & 3.9929 & 27.6000 \\
	QGAC			& $16.20 \pm 0.00$ & 2.6551 & 27.4091 \\
	QGAC-GAN			& $6.93 \pm 0.00$ & 1.8640 & 27.0681 \\
	Bahat			& $13.71 \pm 0.03$ & 0.3598* & 25.4243 \\
	
	Ours        		&  $4.76 \pm 0.01$ & 0.9696 & 25.5758 \\
	Ours-P		& $4.78 \pm 0.01$ & 0.6411* & 25.6008 \\
	
	\rowcolor{gray!25}
	Ground Truth         &  $2.67 \pm 0.00$ & 0 & $\infty$ \\
	\hline
	\end{tabular}
	\label{tab:imagenet}
\end{table}
\endgroup

\begingroup
\newcolumntype{M}[1]{>{\centering\arraybackslash}m{#1}}
\setlength{\tabcolsep}{0pt} %
\renewcommand{\arraystretch}{0} %

\newcommand*\annotatedFigureBoxCustom[3]{\draw[draw=#3, thick] (#1) rectangle (#2);}
\newcommand*\annotatedFigureBox[3]{\annotatedFigureBoxCustom{#1}{#2}{#3}}
\newenvironment {annotatedFigure}[1]
{\centering\begin{tikzpicture}[baseline=(current bounding box.center)]
    \node[anchor=south west, inner sep=0] (image) at (0,0) { #1};
    \begin{scope}[x={(image.south east)},y={(image.north west)}]
}
{
	\end{scope}
	\end{tikzpicture}
}

\begin{figure*}[t]
	\centering
    \begin{tabular}{p{0.025\textwidth} c c c p{0.025\textwidth} c c c}
	    
	    \hspace{3pt}\centered{\begin{turn}{90}\footnotesize{Compressed}\end{turn}}&
	    \noisy[\ffhqw]{5}{ffhq}{03919}&
	    \centered{\adjincludegraphics[height=\ffhqw, trim={{0.15\width} {.65\height} {0.2\width} 0}, clip]{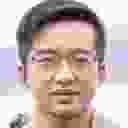}}&
	    \centered{\adjincludegraphics[height=\ffhqw, trim={11px 35px 86px 50px}, clip]{images/ours/5-ffhq/compressed/03919.jpg}}&
	    
	    \hspace{3pt}\centered{\begin{turn}{90}\footnotesize{Ours (1)}\end{turn}}&
	    \ours[\ffhqw]{5}{ffhq}{03919}{2}&
	    \centered{\adjincludegraphics[height=\ffhqw, trim={{0.15\width} {.65\height} {0.2\width} 0}, clip]{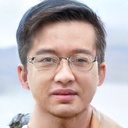}}&
	    \centered{\adjincludegraphics[height=\ffhqw, trim={11px 35px 86px 50px}, clip]{images/ours/5-ffhq/fake_2/03919.jpg}}
	    \\
	    
	    \hspace{3pt}\centered{\begin{turn}{90}\footnotesize{Ours\=/MSE}\end{turn}}&
	    \mmse[\ffhqw]{5}{ffhq}{03919}&
	    \centered{\adjincludegraphics[height=\ffhqw, trim={{0.15\width} {.65\height} {0.2\width} 0}, clip]{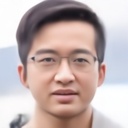}}&
	    \centered{\adjincludegraphics[height=\ffhqw, trim={11px 35px 86px 50px}, clip]{images/mmse/5-ffhq/fake_0/03919.jpg}}&
	    
	    \hspace{3pt}\centered{\begin{turn}{90}\footnotesize{Ours (2)}\end{turn}}&
	    \ours[\ffhqw]{5}{ffhq}{03919}{3}&
	    \centered{\adjincludegraphics[height=\ffhqw, trim={{0.15\width} {.65\height} {0.2\width} 0}, clip]{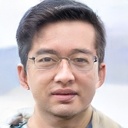}}&
	    \centered{\adjincludegraphics[height=\ffhqw, trim={11px 35px 86px 50px}, clip]{images/ours/5-ffhq/fake_3/03919.jpg}}
	    \\
	    
	    \hspace{3pt}\centered{\begin{turn}{90}\footnotesize{Bahat \etal~\cite{WhatImageExplorable2021bahat}}\end{turn}}&
	    \bahat[\ffhqw]{5}{ffhq}{03919}{0}&
	    \centered{\adjincludegraphics[height=\ffhqw, trim={{0.15\width} {.65\height} {0.2\width} 0}, clip]{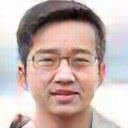}}&
	    \centered{\adjincludegraphics[height=\ffhqw, trim={11px 35px 86px 50px}, clip]{images/bahat/5-ffhq/fake_0/03919.jpg}}&
	    
	    \hspace{3pt}\centered{\begin{turn}{90}\footnotesize{Ours (3)}\end{turn}}&
	    \ours[\ffhqw]{5}{ffhq}{03919}{1}&
	    \centered{\adjincludegraphics[height=\ffhqw, trim={{0.15\width} {.65\height} {0.2\width} 0}, clip]{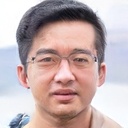}}&
	    \centered{\adjincludegraphics[height=\ffhqw, trim={11px 35px 86px 50px}, clip]{images/ours/5-ffhq/fake_1/03919.jpg}}
	    \\
	    
	    \hspace{3pt}\centered{\begin{turn}{90}\footnotesize{Real}\end{turn}}&
	    \real[\ffhqw]{5}{ffhq}{03919}&
	    \centered{\adjincludegraphics[height=\ffhqw, trim={{0.15\width} {.65\height} {0.2\width} 0}, clip]{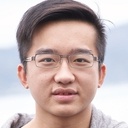}}&
	    \centered{\adjincludegraphics[height=\ffhqw, trim={11px 35px 86px 50px}, clip]{images/ours/5-ffhq/real/03919.jpg}}&
	    
	    \hspace{3pt}\centered{\begin{turn}{90}\footnotesize{Ours (4)}\end{turn}}&
	    \ours[\ffhqw]{5}{ffhq}{03919}{7}&
	    \centered{\adjincludegraphics[height=\ffhqw, trim={{0.15\width} {.65\height} {0.2\width} 0}, clip]{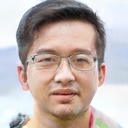}}&
	    \centered{\adjincludegraphics[height=\ffhqw, trim={11px 35px 86px 50px}, clip]{images/ours/5-ffhq/fake_7/03919.jpg}}
	    \\
	    
    \end{tabular}
	\caption{Zoom-in on the results of different recovery methods. Left column: The decompression of a single image from the FFHQ data set compressed using QF=5. Notice the smoothed result of Ours\=/MSE and the artifacts in Bahat's solution, while our method produces sharp and realistic results. Right column: Four realizations from our method that further show the stochastic nature of the results. Note the different hair patterns and ear shapes. %
	}
	\label{fig:ffhq_zoom}
\end{figure*}
\endgroup
\begingroup
\newcolumntype{M}[1]{>{\centering\arraybackslash}m{#1}}
\setlength{\tabcolsep}{0pt} %
\renewcommand{\arraystretch}{0} %

\begin{figure*}[t]
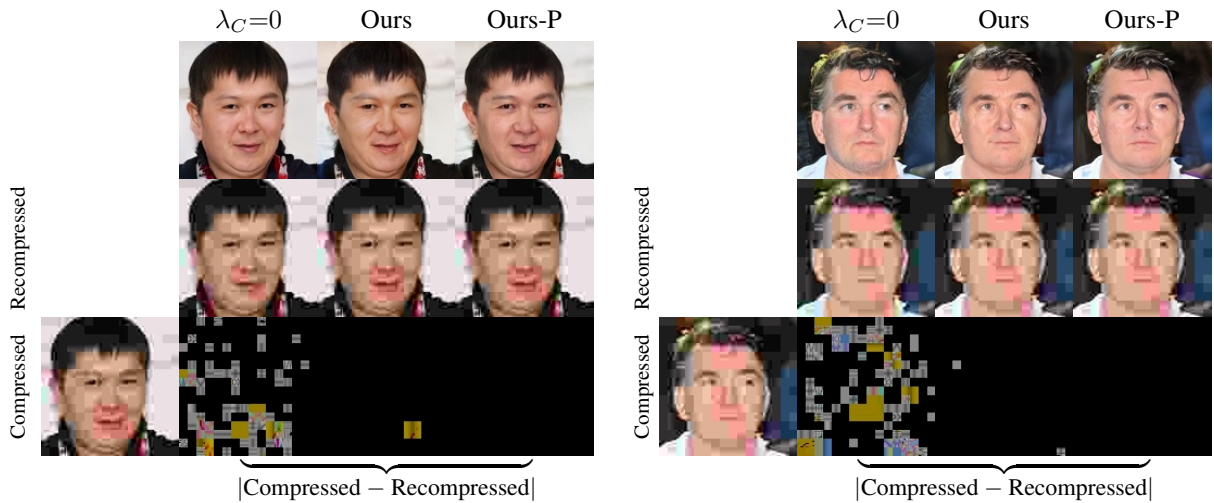

		\centering
	\begin{tabular}{l l}
	    \begin{tabular}{p{0.03\textwidth} c c c c p{0.025\textwidth} }
		    && $\lambda_C{=}0$ & Ours & Ours-P \\
		    
		    &\rule{0pt}{0.8ex}&&\\
	
			&&
			\pscgan{5}{ffhq}{03015}{1}&
		    \ours{5}{ffhq}{03015}{1}&
		    \oursp{5}{ffhq}{03015}{1}
		    \\
		    
		    \hspace{3pt}\centered{\begin{turn}{90}\footnotesize Recompressed\end{turn}}&&
		    \pscganc{5}{ffhq}{03015}{1}&
		    \oursc{5}{ffhq}{03015}{1}&
		    \ourspc{5}{ffhq}{03015}{1}
		    \\
		    
		    \hspace{3pt}\centered{\begin{turn}{90}\footnotesize Compressed\end{turn}}&
		    \noisy{5}{ffhq}{03015}&
		    \pscgand{5}{ffhq}{03015}{1}&
		    \oursd{5}{ffhq}{03015}{1}&
		    \ourspd{5}{ffhq}{03015}{1}
		    \\
		    
		    \rule{0pt}{0.8ex} \\
		    &&\multicolumn{3}{c}{\upbracefill} \\
		    \rule{0pt}{0.8ex} \\
		    &&\multicolumn{3}{c}{\small$|\text{Compressed}-\text{Recompressed}|$}
		    
	    \end{tabular} &
    	\rule{10pt}{0ex}
	    \begin{tabular}{p{0.025\textwidth} c c c c }
		    && $\lambda_C{=}0$ & Ours & Ours-P \\
		    
		    &\rule{0pt}{0.8ex}&&\\
	
			&&
		    \pscgan{5}{ffhq}{03001}{1}&
		    \ours{5}{ffhq}{03001}{1}&
		    \oursp{5}{ffhq}{03001}{1}
		    \\
		    
		    \hspace{3pt}\centered{\begin{turn}{90}\footnotesize Recompressed\end{turn}}&&
		    \pscganc{5}{ffhq}{03001}{1}&
		    \oursc{5}{ffhq}{03001}{1}&
		    \ourspc{5}{ffhq}{03001}{1}
		    \\
		    
		    \hspace{3pt}\centered{\begin{turn}{90}\footnotesize Compressed\end{turn}}&
		    \noisy{5}{ffhq}{03001}&
		    \pscgand{5}{ffhq}{03001}{1}&
		    \oursd{5}{ffhq}{03001}{1}&
		    \ourspd{5}{ffhq}{03001}{1}
		    \\
		    
		    \rule{0pt}{0.8ex} \\
		    &&\multicolumn{3}{c}{\upbracefill} \\
		    \rule{0pt}{0.8ex} \\
		    &&\multicolumn{3}{c}{\small$|\text{Compressed}-\text{Recompressed}|$}
		    
	    \end{tabular}
    \end{tabular}
    
	\caption{The difference between a compressed input and the recompressed outputs of our method with and without explicit consistency penalty and with projection. Values has been rescaled by taking the 4th root for visualization purposes. By adjusting $\lambda_C$ we are able to produce reconstruction with near-perfect consistency (Ours). This allows us to project the results to achieve perfect consistency with minimal impact on the perceptual quality (Ours-P). Quantitative results can be seen in \autoref{tab:ffhq}.}
	\label{fig:ffhq_recompress}
\end{figure*}
\endgroup
\begingroup
\newcolumntype{M}[1]{>{\centering\arraybackslash}m{#1}}
\newcommand{\vcentered}[1]{\begin{tabular}{@{}l@{}} #1 \end{tabular}}
\setlength{\tabcolsep}{0pt} %
\renewcommand{\arraystretch}{0} %

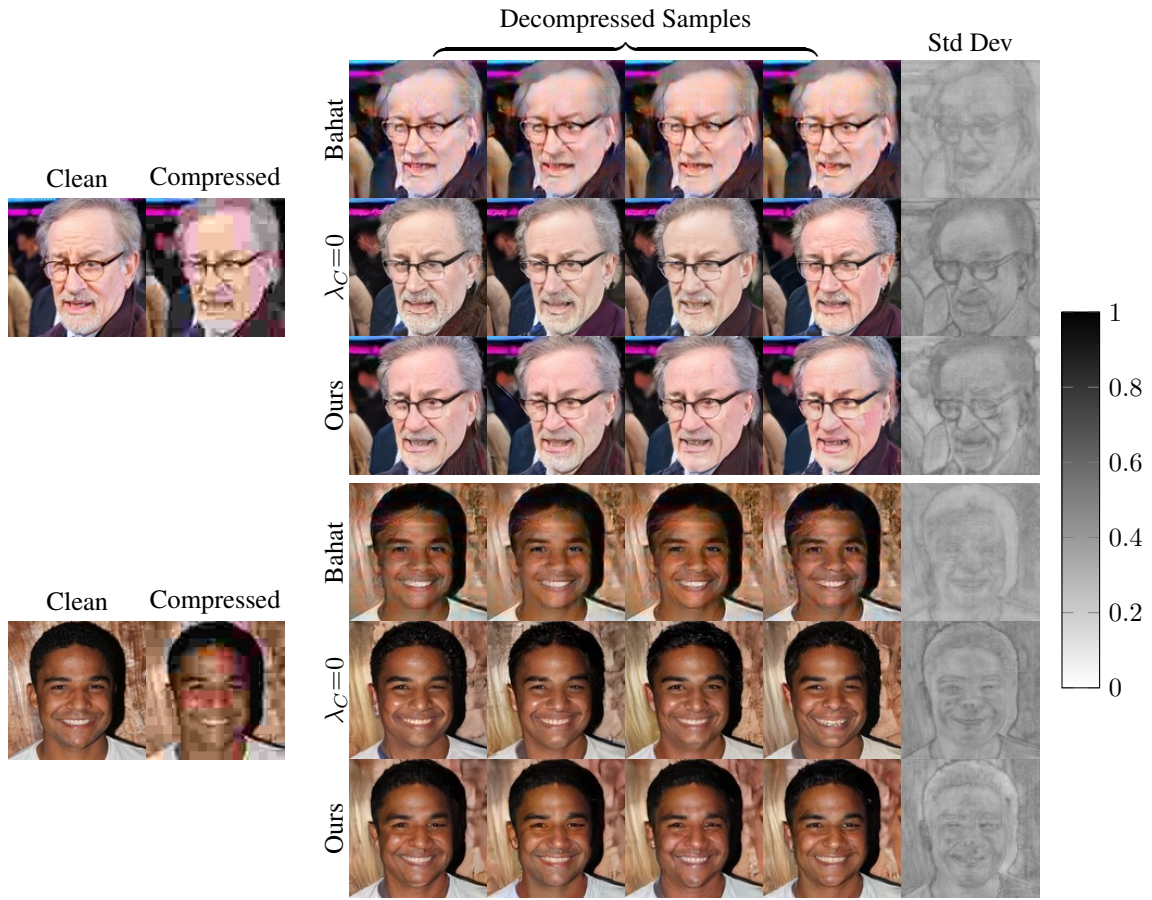
\begin{figure*}[t]
	\centering
	\begin{tabular}{c M{0.05\textwidth}}
    \begin{tabular}{c c @{\hskip 0.025\textwidth} p{0.025\textwidth} c c c c c}
	    &&&\multicolumn{4}{c}{Decompressed Samples}&\\
	    &&&\multicolumn{4}{c}{\downbracefill}&Std Dev\\
	    
	    &\rule{0pt}{0.8ex}&&&&\\

		\begin{tabular}{c}
             \vspace{1.3cm}\\
             Clean
        \end{tabular}&
        \begin{tabular}{c}
             \vspace{1.3cm}\\
             Compressed
        \end{tabular}&
		\hspace{3pt}\centered{\begin{turn}{90}Bahat\end{turn}}&
		\bahat{5}{ffhq}{03121}{0}&
	    \bahat{5}{ffhq}{03121}{1}&
	    \bahat{5}{ffhq}{03121}{2}&
	    \bahat{5}{ffhq}{03121}{3}&
	    \bahats{5}{ffhq}{03121}
		\\
		
		\real{5}{ffhq}{03121}&
		\noisy{5}{ffhq}{03121}&
		\hspace{3pt}\centered{\begin{turn}{90}$\lambda_C{=}0$\end{turn}}&
		\pscgan{5}{ffhq}{03121}{1}&
	    \pscgan{5}{ffhq}{03121}{2}&
	    \pscgan{5}{ffhq}{03121}{3}&
	    \pscgan{5}{ffhq}{03121}{4}&
	    \pscgans{5}{ffhq}{03121}
		\\

		&&
		\hspace{3pt}\centered{\begin{turn}{90}Ours\end{turn}}&
		\ours{5}{ffhq}{03121}{5}&
	    \ours{5}{ffhq}{03121}{2}&
	    \ours{5}{ffhq}{03121}{0}&
	    \ours{5}{ffhq}{03121}{4}&
	    \ourss{5}{ffhq}{03121}
		\\

		\rule{0pt}{0.8ex}\\
		
		\begin{tabular}{c}
             \vspace{1.3cm}\\
             Clean
        \end{tabular}&
        \begin{tabular}{c}
             \vspace{1.3cm}\\
             Compressed
        \end{tabular}&
		\hspace{3pt}\centered{\begin{turn}{90}Bahat\end{turn}}&
		\bahat{5}{ffhq}{03123}{0}&
	    \bahat{5}{ffhq}{03123}{1}&
	    \bahat{5}{ffhq}{03123}{2}&
	    \bahat{5}{ffhq}{03123}{3}&
	    \bahats{5}{ffhq}{03123}
		\\
		
		\real{5}{ffhq}{03123}&
		\noisy{5}{ffhq}{03123}&
		\hspace{3pt}\centered{\begin{turn}{90}$\lambda_C{=}0$\end{turn}}&
		\pscgan{5}{ffhq}{03123}{1}&
	    \pscgan{5}{ffhq}{03123}{2}&
	    \pscgan{5}{ffhq}{03123}{3}&
	    \pscgan{5}{ffhq}{03123}{4}&
	    \pscgans{5}{ffhq}{03123}
		\\

		&&
		\hspace{3pt}\centered{\begin{turn}{90}Ours\end{turn}}&
		\ours{5}{ffhq}{03123}{0}&
	    \ours{5}{ffhq}{03123}{1}&
	    \ours{5}{ffhq}{03123}{7}&
	    \ours{5}{ffhq}{03123}{4}&
	    \ourss{5}{ffhq}{03123}
		\\

    \end{tabular}
    
    &
        \vspace{1.5cm}
        \begin{tikzpicture}
            \begin{axis}[
                hide axis,
                scale only axis,
                height=5cm,
                width=1cm,
                colormap = {whiteblack}{color(0cm)  = (white);color(1cm) = (black)},
                colorbar,
                point meta min=0,
                point meta max=1,
                colorbar style={
                }]
                \addplot [draw=none] coordinates {(0,0)};
            \end{axis}
            \end{tikzpicture}
    
    \end{tabular}
    
	\caption{Stochastic variation of decompressed images using Bahat's and our methods. To the left we present two clean images and their corresponding compressed JPEG version using QF=5. To the right we present 4 realizations using each method, along with per-pixel standard deviation map calculated on 32 samples. For visualization purposes we use the 4th root of the standard deviation and add a color bar (white and black correspond to low and high standard deviations, respectively). All decompressed images were obtained using the default noise injection scheme ($z \sim \mathcal{U}(-1,1)$ for Bahat's method and $z \sim \mathcal{N}(0,I)$ for the others).}
	\label{fig:ffhq_variation}
\end{figure*}
\endgroup
\begingroup
\newcolumntype{M}[1]{>{\centering\arraybackslash}m{#1}}
\newcommand{\vcentered}[1]{\begin{tabular}{@{}l@{}} #1 \end{tabular}}
\setlength{\tabcolsep}{0pt} %
\renewcommand{\arraystretch}{0} %

\begin{figure*}[htpb]
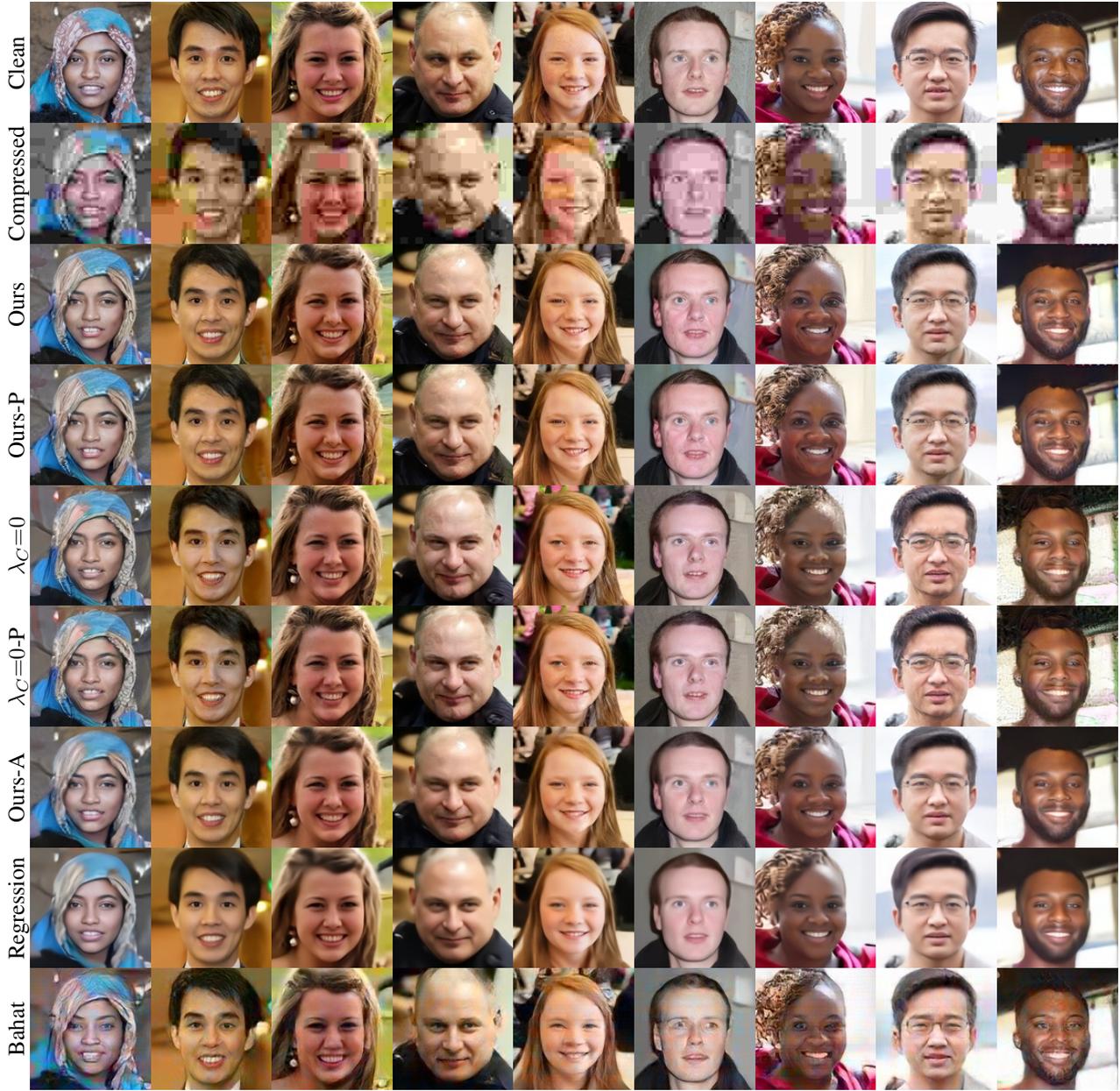

	\centering
    \begin{tabular}{p{0.025\textwidth} c c c c c c c c c}
	    
	    &\rule{0pt}{0.8ex}&&&&\\

		\hspace{3pt}\centered{\begin{turn}{90}Clean\end{turn}}&
		\real{5}{ffhq}{03003}&
	    \real{5}{ffhq}{03007}&
	    \real{5}{ffhq}{03621}&
	    \real{5}{ffhq}{03649}&
	    \real{5}{ffhq}{03844}&
	    \real{5}{ffhq}{03876}&
	    \real{5}{ffhq}{03910}&
	    \real{5}{ffhq}{03919}&
	    \real{5}{ffhq}{03942}
		\\
		
		\hspace{3pt}\centered{\begin{turn}{90}Compressed\end{turn}}&
		\noisy{5}{ffhq}{03003}&
	    \noisy{5}{ffhq}{03007}&
	    \noisy{5}{ffhq}{03621}&
	    \noisy{5}{ffhq}{03649}&
	    \noisy{5}{ffhq}{03844}&
	    \noisy{5}{ffhq}{03876}&
	    \noisy{5}{ffhq}{03910}&
	    \noisy{5}{ffhq}{03919}&
	    \noisy{5}{ffhq}{03942}
		\\
		
		\hspace{3pt}\centered{\begin{turn}{90}Ours\end{turn}}&
		\ours{5}{ffhq}{03003}{0}&
	    \ours{5}{ffhq}{03007}{0}&
	    \ours{5}{ffhq}{03621}{0}&
	    \ours{5}{ffhq}{03649}{0}&
	    \ours{5}{ffhq}{03844}{0}&
	    \ours{5}{ffhq}{03876}{0}&
	    \ours{5}{ffhq}{03910}{0}&
	    \ours{5}{ffhq}{03919}{0}&
	    \ours{5}{ffhq}{03942}{0}
		\\
		
		\hspace{3pt}\centered{\begin{turn}{90}Ours-P\end{turn}}&
		\oursp{5}{ffhq}{03003}{0}&
	    \oursp{5}{ffhq}{03007}{0}&
	    \oursp{5}{ffhq}{03621}{0}&
	    \oursp{5}{ffhq}{03649}{0}&
	    \oursp{5}{ffhq}{03844}{0}&
	    \oursp{5}{ffhq}{03876}{0}&
	    \oursp{5}{ffhq}{03910}{0}&
	    \oursp{5}{ffhq}{03919}{0}&
	    \oursp{5}{ffhq}{03942}{0}
		\\
		
		\hspace{3pt}\centered{\begin{turn}{90}$\lambda_C{=}0$\end{turn}}&
		\pscgan{5}{ffhq}{03003}{0}&
	    \pscgan{5}{ffhq}{03007}{0}&
	    \pscgan{5}{ffhq}{03621}{0}&
	    \pscgan{5}{ffhq}{03649}{0}&
	    \pscgan{5}{ffhq}{03844}{0}&
	    \pscgan{5}{ffhq}{03876}{0}&
	    \pscgan{5}{ffhq}{03910}{0}&
	    \pscgan{5}{ffhq}{03919}{0}&
	    \pscgan{5}{ffhq}{03942}{0}
		\\
		
		\hspace{3pt}\centered{\begin{turn}{90}$\lambda_C{=}0$-P\end{turn}}&
		\pscganp{5}{ffhq}{03003}{0}&
	    \pscganp{5}{ffhq}{03007}{0}&
	    \pscganp{5}{ffhq}{03621}{0}&
	    \pscganp{5}{ffhq}{03649}{0}&
	    \pscganp{5}{ffhq}{03844}{0}&
	    \pscganp{5}{ffhq}{03876}{0}&
	    \pscganp{5}{ffhq}{03910}{0}&
	    \pscganp{5}{ffhq}{03919}{0}&
	    \pscganp{5}{ffhq}{03942}{0}
		\\
		
		\hspace{3pt}\centered{\begin{turn}{90}Ours-A\end{turn}}&
		\oursa{5}{ffhq}{03003}&
	    \oursa{5}{ffhq}{03007}&
	    \oursa{5}{ffhq}{03621}&
	    \oursa{5}{ffhq}{03649}&
	    \oursa{5}{ffhq}{03844}&
	    \oursa{5}{ffhq}{03876}&
	    \oursa{5}{ffhq}{03910}&
	    \oursa{5}{ffhq}{03919}&
	    \oursa{5}{ffhq}{03942}
		\\
		
		\hspace{3pt}\centered{\begin{turn}{90}Regression\end{turn}}&
		\mmse{5}{ffhq}{03003}&
	    \mmse{5}{ffhq}{03007}&
	    \mmse{5}{ffhq}{03621}&
	    \mmse{5}{ffhq}{03649}&
	    \mmse{5}{ffhq}{03844}&
	    \mmse{5}{ffhq}{03876}&
	    \mmse{5}{ffhq}{03910}&
	    \mmse{5}{ffhq}{03919}&
	    \mmse{5}{ffhq}{03942}
		\\
		
		\hspace{3pt}\centered{\begin{turn}{90}Bahat\end{turn}}&
		\bahat{5}{ffhq}{03003}{0}&
	    \bahat{5}{ffhq}{03007}{0}&
	    \bahat{5}{ffhq}{03621}{0}&
	    \bahat{5}{ffhq}{03649}{0}&
	    \bahat{5}{ffhq}{03844}{0}&
	    \bahat{5}{ffhq}{03876}{0}&
	    \bahat{5}{ffhq}{03910}{0}&
	    \bahat{5}{ffhq}{03919}{0}&
	    \bahat{5}{ffhq}{03942}{0}
		\\

    \end{tabular}
    
	\caption{Decompression results using different methods on FFHQ images compressed using JPEG with QF=5. For stochastic methods (all except for Regression), the default noise injection scheme ($z \sim \mathcal{U}(-1,1)$ for Bahat's method and $z \sim \mathcal{N}(0,I)$ for the others) was used during inference.}
	\label{fig:ffhq_perceptual}
\end{figure*}
\endgroup

\begingroup
\newcolumntype{M}[1]{>{\centering\arraybackslash}m{#1}}
\newcommand{\vcentered}[1]{\begin{tabular}{@{}l@{}} #1 \end{tabular}}
\setlength{\tabcolsep}{0pt} %
\renewcommand{\arraystretch}{0} %

\setlength{\hero}{0.18\textwidth}

\begin{figure*}[htpb]
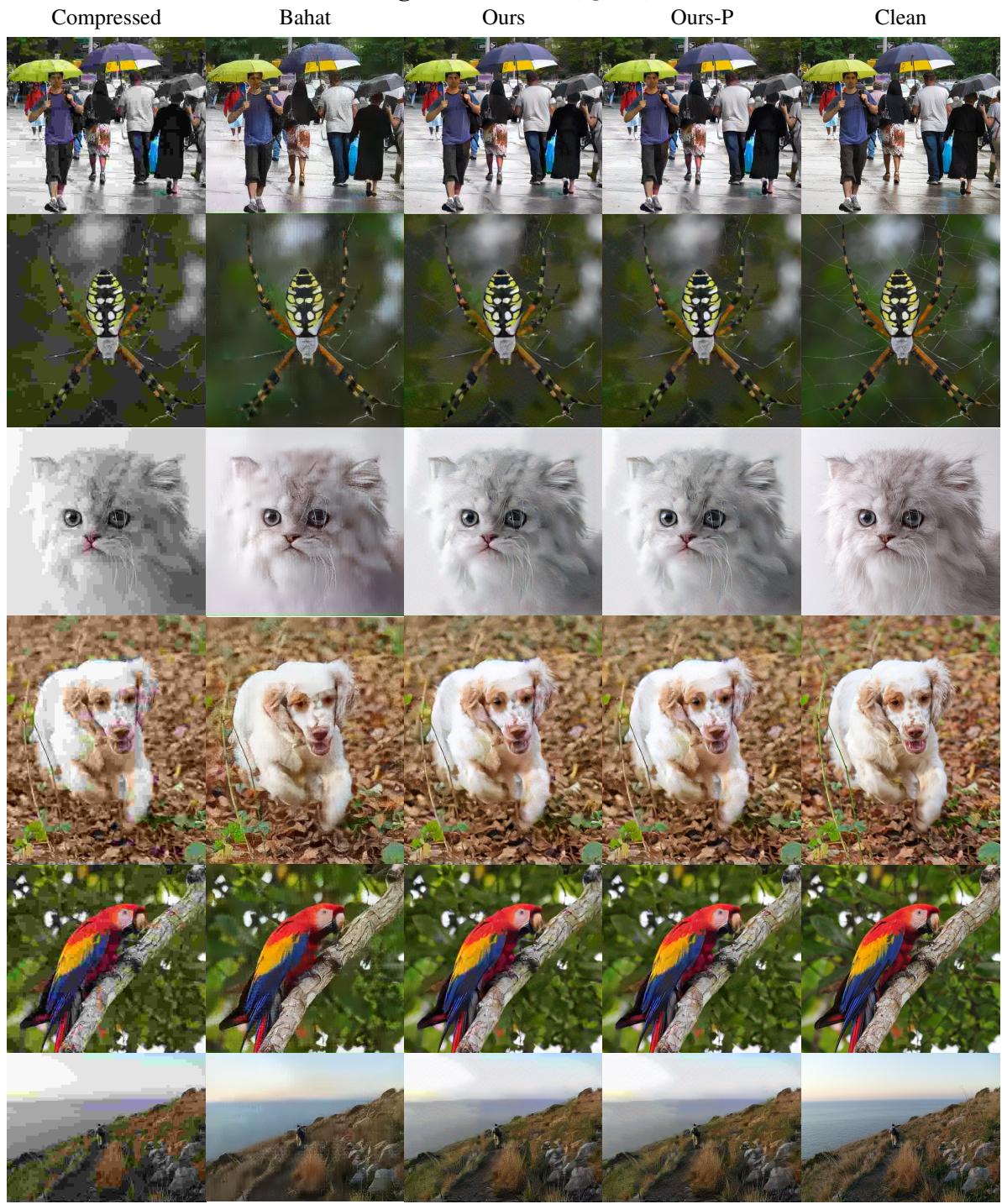

	\centering
    \begin{tabular}{c c c c c}
    
    	\multicolumn{5}{c}{\large{\textbf{ImageNet-ctest10k (QF=5)}}}\\
    	
    	\rule{0pt}{0.8ex}\\
	    
	    Compressed &
	    Bahat &
	    Ours &
	    Ours-P &
	    Clean
	    \\
	    
	    \rule{0pt}{0.8ex}\\

		\noisy[\hero]{5}{imagenet}{ILSVRC2012_val_00040204}&
	    \bahat[\hero]{5}{imagenet}{ILSVRC2012_val_00040204}{0}&
	    \ours[\hero]{5}{imagenet}{ILSVRC2012_val_00040204}{0}&
	    \oursp[\hero]{5}{imagenet}{ILSVRC2012_val_00040204}{0}&
	    \real[\hero]{5}{imagenet}{ILSVRC2012_val_00040204}
		\\
		
		\noisy[\hero]{5}{imagenet}{ILSVRC2012_val_00040241}&
	    \bahat[\hero]{5}{imagenet}{ILSVRC2012_val_00040241}{0}&
	    \ours[\hero]{5}{imagenet}{ILSVRC2012_val_00040241}{0}&
	    \oursp[\hero]{5}{imagenet}{ILSVRC2012_val_00040241}{0}&
	    \real[\hero]{5}{imagenet}{ILSVRC2012_val_00040241}
		\\
		
		\noisy[\hero]{5}{imagenet}{ILSVRC2012_val_00037733}&
	    \bahat[\hero]{5}{imagenet}{ILSVRC2012_val_00037733}{0}&
	    \ours[\hero]{5}{imagenet}{ILSVRC2012_val_00037733}{0}&
	    \oursp[\hero]{5}{imagenet}{ILSVRC2012_val_00037733}{0}&
	    \real[\hero]{5}{imagenet}{ILSVRC2012_val_00037733}
		\\
		
		\noisy[\hero]{5}{imagenet}{ILSVRC2012_val_00040291}&
	    \bahat[\hero]{5}{imagenet}{ILSVRC2012_val_00040291}{0}&
	    \ours[\hero]{5}{imagenet}{ILSVRC2012_val_00040291}{0}&
	    \oursp[\hero]{5}{imagenet}{ILSVRC2012_val_00040291}{0}&
	    \real[\hero]{5}{imagenet}{ILSVRC2012_val_00040291}
		\\
		
		\noisy[\hero]{5}{imagenet}{ILSVRC2012_val_00040359}&
	    \bahat[\hero]{5}{imagenet}{ILSVRC2012_val_00040359}{0}&
	    \ours[\hero]{5}{imagenet}{ILSVRC2012_val_00040359}{0}&
	    \oursp[\hero]{5}{imagenet}{ILSVRC2012_val_00040359}{0}&
	    \real[\hero]{5}{imagenet}{ILSVRC2012_val_00040359}
		\\
		
		\noisy[\hero]{5}{imagenet}{ILSVRC2012_val_00042182}&
	    \bahat[\hero]{5}{imagenet}{ILSVRC2012_val_00042182}{0}&
	    \ours[\hero]{5}{imagenet}{ILSVRC2012_val_00042182}{0}&
	    \oursp[\hero]{5}{imagenet}{ILSVRC2012_val_00042182}{0}&
	    \real[\hero]{5}{imagenet}{ILSVRC2012_val_00042182}
		\\

    \end{tabular}
    
	\caption{Decompression results using our method and Bahat~\etal's on ImageNet JPEG compressed images with QF=5.}
	\label{fig:imagenet_5_perceptual}
\end{figure*}
\endgroup

\begingroup
\newcolumntype{M}[1]{>{\centering\arraybackslash}m{#1}}
\newcommand{\vcentered}[1]{\begin{tabular}{@{}l@{}} #1 \end{tabular}}
\setlength{\tabcolsep}{0pt} %
\renewcommand{\arraystretch}{0} %

\setlength{\hero}{0.175\textwidth}

\begin{figure*}[htpb]
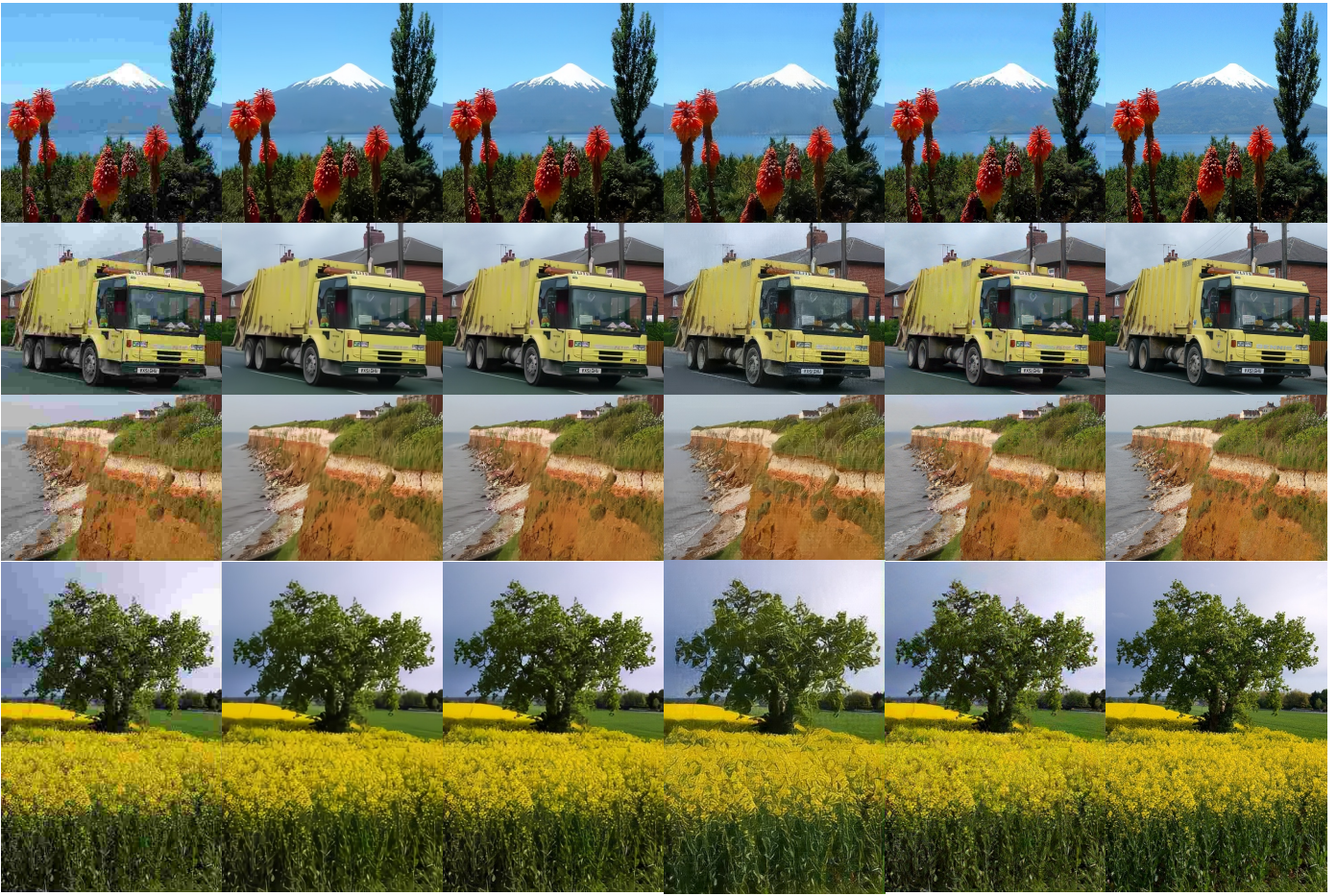

	\centering
    \begin{tabular}{c c c c c c}
    
    	\multicolumn{6}{c}{\large{\textbf{ImageNet-ctest10k (QF=10)}}}\\
    	
    	\rule{0pt}{0.8ex}\\
	    
	    Compressed &
	    QGAC &
	    QGAC-GAN &
	    Bahat &
	    Ours-P &
	    Clean
	    \\
	    
	    \rule{0pt}{0.8ex}\\

		\noisy[\hero]{10}{imagenet}{ILSVRC2012_val_00039999}&
		\qgac[\hero]{10}{imagenet}{ILSVRC2012_val_00039999}&
		\qgacg[\hero]{10}{imagenet}{ILSVRC2012_val_00039999}&
	    \bahat[\hero]{10}{imagenet}{ILSVRC2012_val_00039999}{0}&
	    \oursp[\hero]{10}{imagenet}{ILSVRC2012_val_00039999}{0}&
	    \real[\hero]{10}{imagenet}{ILSVRC2012_val_00039999}
		\\
		
		\noisy[\hero]{10}{imagenet}{ILSVRC2012_val_00041478}&
		\qgac[\hero]{10}{imagenet}{ILSVRC2012_val_00041478}&
		\qgacg[\hero]{10}{imagenet}{ILSVRC2012_val_00041478}&
	    \bahat[\hero]{10}{imagenet}{ILSVRC2012_val_00041478}{0}&
	    \oursp[\hero]{10}{imagenet}{ILSVRC2012_val_00041478}{0}&
	    \real[\hero]{10}{imagenet}{ILSVRC2012_val_00041478}
		\\

		\noisy[\hero]{10}{imagenet}{ILSVRC2012_val_00044389}&
		\qgac[\hero]{10}{imagenet}{ILSVRC2012_val_00044389}&
		\qgacg[\hero]{10}{imagenet}{ILSVRC2012_val_00044389}&
	    \bahat[\hero]{10}{imagenet}{ILSVRC2012_val_00044389}{0}&
	    \oursp[\hero]{10}{imagenet}{ILSVRC2012_val_00044389}{0}&
	    \real[\hero]{10}{imagenet}{ILSVRC2012_val_00044389}
		\\
		
		\noisy[\hero]{10}{imagenet}{ILSVRC2012_val_00040363}&
		\qgac[\hero]{10}{imagenet}{ILSVRC2012_val_00040363}&
		\qgacg[\hero]{10}{imagenet}{ILSVRC2012_val_00040363}&
	    \bahat[\hero]{10}{imagenet}{ILSVRC2012_val_00040363}{0}&
	    \oursp[\hero]{10}{imagenet}{ILSVRC2012_val_00040363}{0}&
	    \real[\hero]{10}{imagenet}{ILSVRC2012_val_00040363}
		\\

    \end{tabular}
    
	\caption{Decompression results using different methods on ImageNet JPEG compressed images with QF=10.}
	\label{fig:imagenet_10_perceptual}
\end{figure*}
\endgroup
\begingroup
\newcolumntype{M}[1]{>{\centering\arraybackslash}m{#1}}
\newcommand{\vcentered}[1]{\begin{tabular}{@{}l@{}} #1 \end{tabular}}
\setlength{\tabcolsep}{0pt} %
\renewcommand{\arraystretch}{0} %

\setlength{\hero}{0.175\textwidth}

\newcommand{\addimgcol}[7][1]{
	\centered{
	   	\begin{tikzpicture}[
	   		baseline=-2.45,
	   		spy using outlines={magnification=#3, circle, height=1.5cm, width=1.5cm, yellow, every spy on node/.append style={thick}, connect spies},
	   		]
			\node[inner sep=0pt]{\scalebox{#1}[1]{\adjincludegraphics[width=\hero, trim={#4}, clip]{images/ours/#7-live1/compressed/#2}}};
			\spy on (#5) in node at (#6);
		\end{tikzpicture}}&
		
		\centered{
		\begin{tikzpicture}[
	   		baseline=-2.45,
	   		spy using outlines={magnification=#3, circle, height=1.5cm, width=1.5cm, yellow, every spy on node/.append style={thick}, connect spies},
	   		]
			\node[inner sep=0pt]{\scalebox{#1}[1]{\adjincludegraphics[width=\hero, trim={#4}, clip]{images/qgac/#7-live1/#2}}};
			\spy on (#5) in node at (#6);
		\end{tikzpicture}}&
		
		\centered{
		\begin{tikzpicture}[
	   		baseline=-2.45,
	   		spy using outlines={magnification=#3, circle, height=1.5cm, width=1.5cm, yellow, every spy on node/.append style={thick}, connect spies},
	   		]
			\node[inner sep=0pt]{\scalebox{#1}[1]{\adjincludegraphics[width=\hero, trim={#4}, clip]{images/qgac-gan/#7-live1/#2}}};
			\spy on (#5) in node at (#6);
		\end{tikzpicture}}&
		
		\centered{
	   	\begin{tikzpicture}[
	   		baseline=-2.45,
	   		spy using outlines={magnification=#3, circle, height=1.5cm, width=1.5cm, yellow, every spy on node/.append style={thick}, connect spies},
	   		]
			\node[inner sep=0pt]{\scalebox{#1}[1]{\adjincludegraphics[width=\hero, trim={#4}, clip]{images/bahat/#7-live1/fake_0/#2}}};
			\spy on (#5) in node at (#6);
		\end{tikzpicture}}&

		\centered{
		\begin{tikzpicture}[
	   		baseline=-2.45,
	   		spy using outlines={magnification=#3, circle, height=1.5cm, width=1.5cm, yellow, every spy on node/.append style={thick}, connect spies},
	   		]
			\node[inner sep=0pt]{\scalebox{#1}[1]{\adjincludegraphics[width=\hero, trim={#4}, clip]{images/ours-p/#7-live1/fake_0/#2}}};
			\spy on (#5) in node at (#6);
		\end{tikzpicture}}&
		
		\centered{
		\begin{tikzpicture}[
	   		baseline=-2.45,
	   		spy using outlines={magnification=#3, circle, height=1.5cm, width=1.5cm, yellow, every spy on node/.append style={thick}, connect spies},
	   		]
			\node[inner sep=0pt]{\scalebox{#1}[1]{\adjincludegraphics[width=\hero, trim={#4}, clip]{images/ours/#7-live1/real/#2}}};
			\spy on (#5) in node at (#6);
		\end{tikzpicture}}
		
	    \\
}

\begin{figure*}[t]
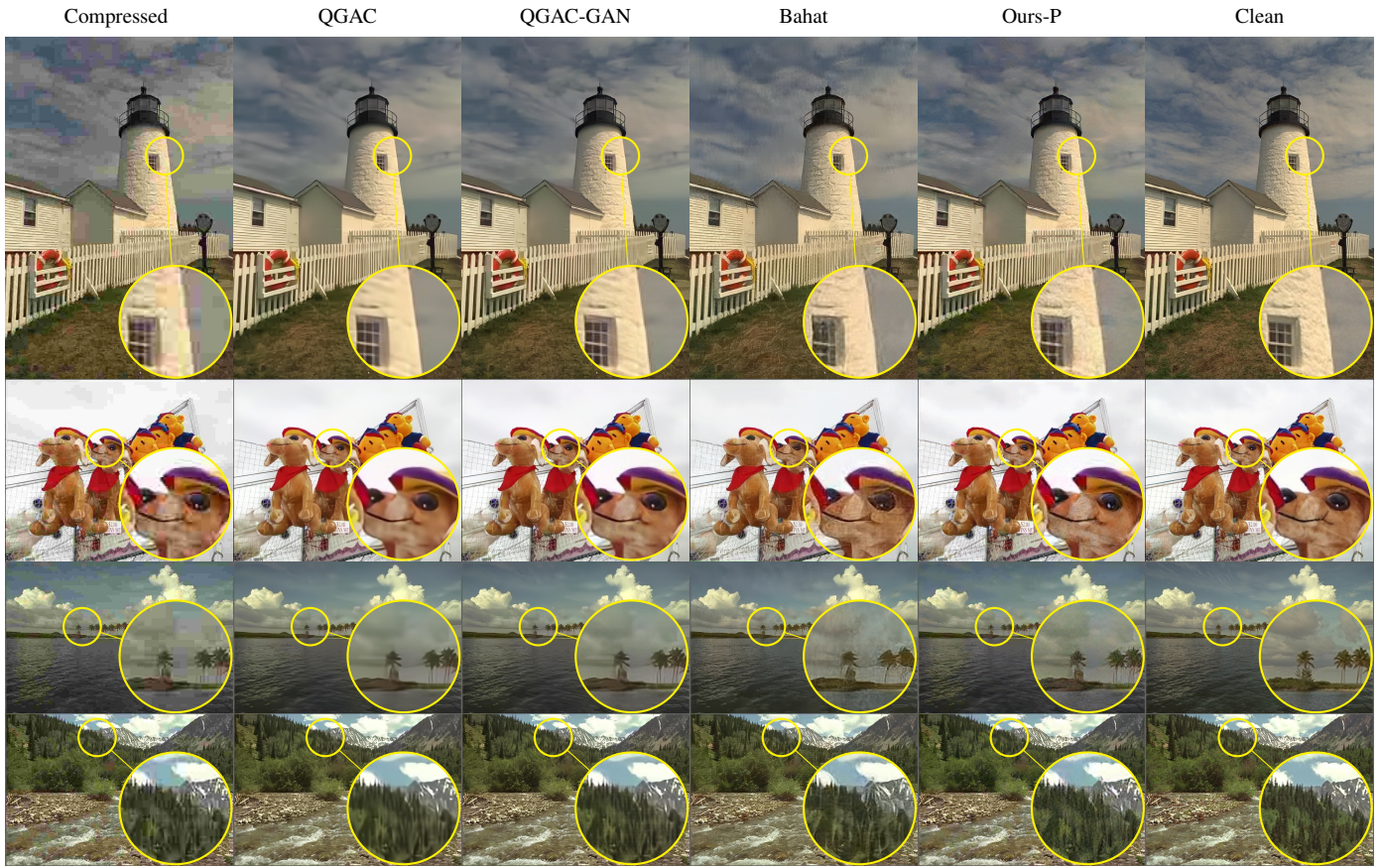

	\centering
    \begin{tabular}{c c c c c c}
    
	    \multicolumn{6}{c}{\large{\textbf{LIVE1 (QF=10)}}}\\
    	
    	\rule{0pt}{0.8ex}\\

	    \footnotesize{Compressed} &
	    \footnotesize{QGAC} &
	    \footnotesize{QGAC-GAN} &
	    \footnotesize{Bahat} &
	    \footnotesize{Ours-P} &
	    \footnotesize{Clean}
	    \\
	    
	   	\rule{0pt}{0.8ex}\\
	   	
	   	\addimgcol{lighthouse}{3}{0 0 0 0}{0.6,0.7}{0.75,-1.525}{10}
	   	\addimgcol{carnivaldolls}{3}{0 0 0 0}{-0.2,0.3}{0.75,-0.45}{10}
	   	\addimgcol{ocean}{3}{0 0 0 0}{-0.5,0.15}{0.75,-0.25}{10}
	   	\addimgcol{stream}{3}{0 0 0 0}{-0.3,0.7}{0.75,-0.25}{10}

    \end{tabular}
    \caption{Decompression results using different methods on LIVE1 JPEG compressed images with QF=10.}
	\label{fig:live1_perceptual}
\end{figure*}
\endgroup
\begingroup
\newcolumntype{M}[1]{>{\centering\arraybackslash}m{#1}}
\newcommand{\vcentered}[1]{\begin{tabular}{@{}l@{}} #1 \end{tabular}}
\setlength{\tabcolsep}{0pt} %
\renewcommand{\arraystretch}{0} %

\setlength{\hero}{0.175\textwidth}

\newcommand{\addimgcol}[8][1]{
	\centered{
	   	\begin{tikzpicture}[
	   		baseline=-2.45,
	   		spy using outlines={magnification=#3, circle, height=#8, width=#8, yellow, every spy on node/.append style={thick}, connect spies},
	   		]
			\node[inner sep=0pt]{\scalebox{#1}[1]{\adjincludegraphics[width=\hero, trim={#4}, clip]{images/ours/#7-bsds/compressed/#2}}};
			\spy on (#5) in node at (#6);
		\end{tikzpicture}}&
		
		\centered{
		\begin{tikzpicture}[
	   		baseline=-2.45,
	   		spy using outlines={magnification=#3, circle, height=#8, width=#8, yellow, every spy on node/.append style={thick}, connect spies},
	   		]
			\node[inner sep=0pt]{\scalebox{#1}[1]{\adjincludegraphics[width=\hero, trim={#4}, clip]{images/qgac/#7-bsds/#2}}};
			\spy on (#5) in node at (#6);
		\end{tikzpicture}}&
		
		\centered{
		\begin{tikzpicture}[
	   		baseline=-2.45,
	   		spy using outlines={magnification=#3, circle, height=#8, width=#8, yellow, every spy on node/.append style={thick}, connect spies},
	   		]
			\node[inner sep=0pt]{\scalebox{#1}[1]{\adjincludegraphics[width=\hero, trim={#4}, clip]{images/qgac-gan/#7-bsds/#2}}};
			\spy on (#5) in node at (#6);
		\end{tikzpicture}}&
		
		\centered{
	   	\begin{tikzpicture}[
	   		baseline=-2.45,
	   		spy using outlines={magnification=#3, circle, height=#8, width=#8, yellow, every spy on node/.append style={thick}, connect spies},
	   		]
			\node[inner sep=0pt]{\scalebox{#1}[1]{\adjincludegraphics[width=\hero, trim={#4}, clip]{images/bahat/#7-bsds/fake_0/#2}}};
			\spy on (#5) in node at (#6);
		\end{tikzpicture}}&

		\centered{
		\begin{tikzpicture}[
	   		baseline=-2.45,
	   		spy using outlines={magnification=#3, circle, height=#8, width=#8, yellow, every spy on node/.append style={thick}, connect spies},
	   		]
			\node[inner sep=0pt]{\scalebox{#1}[1]{\adjincludegraphics[width=\hero, trim={#4}, clip]{images/ours-p/#7-bsds/fake_0/#2}}};
			\spy on (#5) in node at (#6);
		\end{tikzpicture}}&
		
		\centered{
		\begin{tikzpicture}[
	   		baseline=-2.45,
	   		spy using outlines={magnification=#3, circle, height=#8, width=#8, yellow, every spy on node/.append style={thick}, connect spies},
	   		]
			\node[inner sep=0pt]{\scalebox{#1}[1]{\adjincludegraphics[width=\hero, trim={#4}, clip]{images/ours/#7-bsds/real/#2}}};
			\spy on (#5) in node at (#6);
		\end{tikzpicture}}
		
	    \\
}

\begin{figure*}[t]
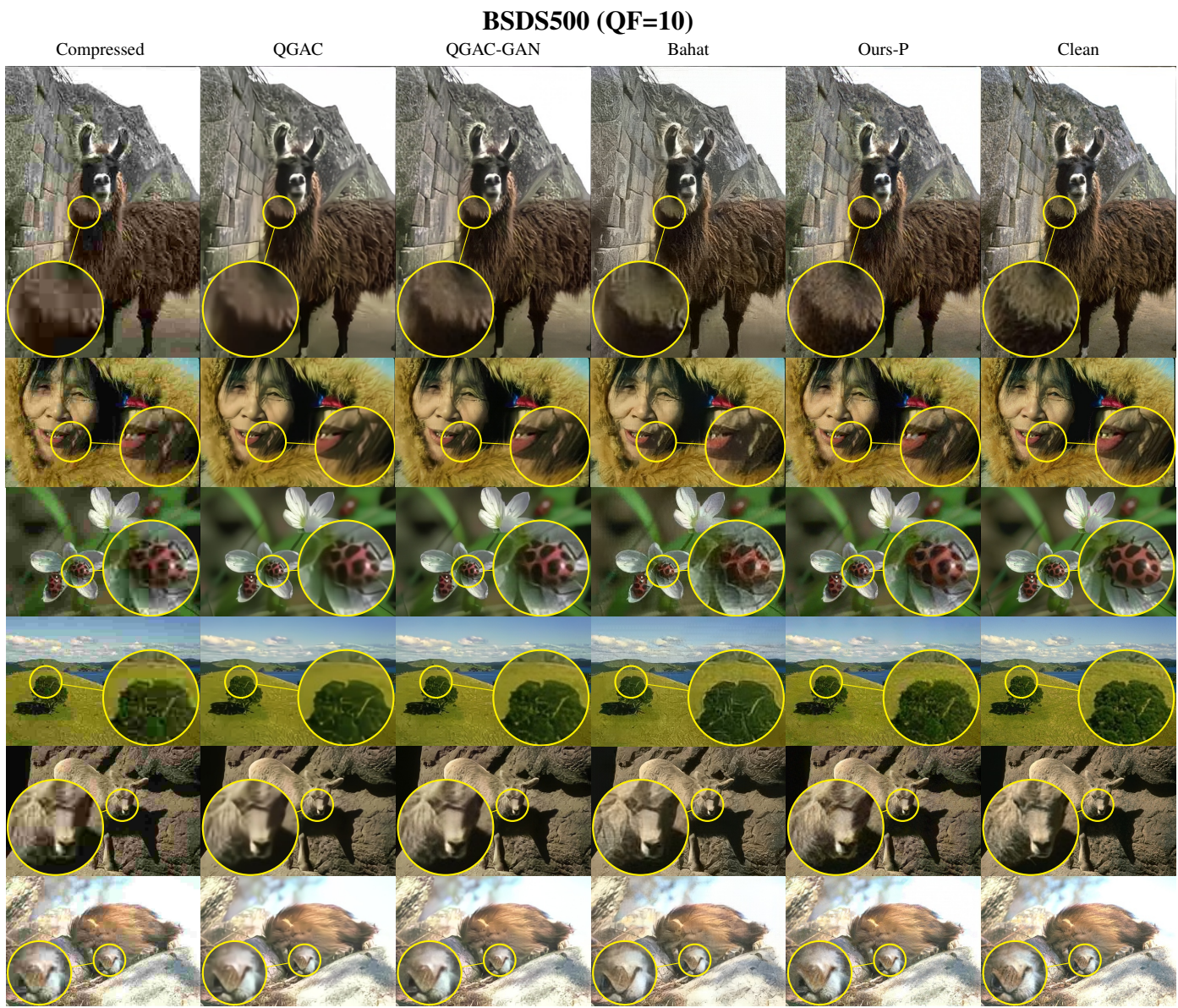

	\centering
    \begin{tabular}{c c c c c c}
    
	    \multicolumn{6}{c}{\large{\textbf{BSDS500 (QF=10)}}}\\
    	
    	\rule{0pt}{0.8ex}\\

	    \footnotesize{Compressed} &
	    \footnotesize{QGAC} &
	    \footnotesize{QGAC-GAN} &
	    \footnotesize{Bahat} &
	    \footnotesize{Ours-P} &
	    \footnotesize{Clean}
	    \\
	    
	   	\rule{0pt}{0.8ex}\\
	   	
	   	\addimgcol{6046}{3}{0 0 0 0}{-0.3,0.0}{-0.75,-1.525}{10}{1.5cm}
	   	\addimgcol{14085}{2}{0 0 0 0}{-0.5,-0.3}{0.9,-0.375}{10}{1.25cm}
	   	\addimgcol{35008}{3}{0 0 0 0}{-0.4,-0.3}{0.75,-0.25}{10}{1.5cm}
	   	\addimgcol{36046}{3}{0 0 0 0}{-0.9,0}{0.75,-0.25}{10}{1.5cm}
	   	\addimgcol{41025}{3}{0 0 0 0}{0.3,0.1}{-0.75,-0.25}{10}{1.5cm}
	   	\addimgcol{105025}{2}{0 0 0 0}{0.1,-0.3}{-1,-0.5}{10}{1cm}

    \end{tabular}
    \caption{Decompression results using different methods on BSDS500 JPEG compressed images with QF=10.}
	\label{fig:bsds_perceptual}
\end{figure*}
\endgroup

\end{document}